\documentclass[preprint,12pt]{revtex4}

\usepackage{amssymb}
\usepackage{amsmath,graphicx}
\usepackage{dsfont,color}

\newcommand{\rd}{{\rm d}}
\newcommand{\ri}{{\rm i}}
\newcommand{\rs}{{\rm s}}
\newcommand{\rp}{{\rm p}}

\newcommand{\ra}{{\rm a}}

\newcommand{\kb}{k_{\rm B}}

\renewcommand{\Im}{{\rm Im}}
\renewcommand{\Re}{{\rm Re}}

\newcommand{\GE}{{\mathds{G}^{\rm E}}}
\newcommand{\GH}{{\mathds{G}^{\rm H}}}

\begin{document}

%\begin{frontmatter}



\title{Radiative heat flux predictions in hyperbolic metamaterials}


\author{Maria Tschikin,$^1$ Svend-Age Biehs,$^1$ Philippe Ben-Abdallah,$^2$ Slawa Lang,$^{3}$ Alexander Yu. Petrov,$^3$ and Manfred Eich$^3$}

\address{$^1$Institut f\"{u}r Physik, Carl von Ossietzky Universit\"{a}t, D-26111 Oldenburg, Germany	\\
				 $^2$Laboratoire Charles Fabry, UMR 8501, Institut d'Optique, CNRS, Universit\'{e} Paris-Sud 11, 2, Avenue
Augustin Fresnel, 91127 Palaiseau Cedex, France	\\
				 $^3$Institute of Optical and Electronic Materials, Hamburg University of Technology, 21073 Hamburg, Germany}

%\fntext[label2]{E-Mail: biehs@theorie.physik.uni-oldenburg.de; Telephone: +49 441 798 3069}
%\tonoteref{biehs@theorie.physik.uni-oldenburg.de}
%\phone{Telephone: +49 441 798 3069}

\begin{abstract}
The transport of heat mediated by thermal photons in hyperbolic multilayer metamaterials is studied using the fluctuational electrodynamics theory. We discuss the dependence of the attenuation length and the heat flux on the design parameters of the multilayer structure. We demonstrate that in comparison to bulk materials the flux inside layered hyperbolic materials can be transported at much longer distances, making these media very promising for thermal management and for near-field energy harvesting. 
\end{abstract}

%\begin{keyword}
%hyperbolic metamaterials \sep near-field radiative heat transfer \sep penetration depth \sep thermophotovoltaics


%\end{keyword}

\maketitle

%\end{frontmatter}


\section{Introduction}
%%%%%%%%%%%%%%%%%%%%%%%%%%%%%%%%%%%%%%%%%%%%%%%%%%%%%%%%%%%%

Nanoscale radiative heat transfer has attracted a lot of attention in the last few years because of Polder and van Hove's prediction~\cite{PvH1971} on the possibility to observe heat fluxes at subwavelength distances which are several orders of magnitude larger than those obtained by the blackbody theory. Recent experimental results~\cite{KittelEtAlPRL2005,NanolettArvind,NatureEmmanuel,HuEtAl2008,ShenEtAl2009,ShiEtAl2013,Ottens2011,Kralik2012} have confirmed these theoretical predictions~\cite{PvH1971,Otey2011,KruegerEtAl2011}.

This increased radiative heat transfer in the near-field regime might be used for different applications as for example near-field imaging~\cite{Yannick,KittelEtAl2008,HuthEtAl2011,WorbesEtAl2013}, nanoscale thermal management by heat flux rectification, amplification and storage~\cite{OteyEtAl2010,IizukaEtAl2012,HuangEtAl2013,PBASAB_PRL2014,PBASABDiode2013,Nefzaoui2013,KubitskyiEtAl2014,DyakovEtAl2014}, and near-field thermophotovoltaics~\cite{MatteoEtAl2001,NarayanaswamyChen2003,LarocheEtAl06,ParkEtAl2007,ZhangReview,GuoEtAl2013}. In particular, for near-field thermophotovoltaic (nTPV) applications it is desirable to have large heat fluxes which are quasi-monochromatic at the bandgap frequency of the thermophotovoltaic cell. Now, it could be shown theoretically that for phonon-polaritonic materials the heat flux is quasi-monochromatic at the surface phonon-frequency resulting in heat fluxes which can be orders of magnitude larger than the blackbody result~\cite{JoulainEtAl2005} due to the large number of contributing surface modes~\cite{BiehsEtAl2010}. This is the reason why phonon-polaritonic media are used in most experiments~\cite{NanolettArvind,NatureEmmanuel,HuEtAl2008,ShenEtAl2009,Ottens2011}. However, it should be kept in mind that there are also upper limits for this surface mode contribution as shown in ~\cite{PBAjoulain2010,Volok2004,BasuZhanglim2009,WangEtAl2009}.

On the other hand, the nanoscale heat flux between two halfspaces separated by a distance $d$ which is due to surface modes is absorbed on a very thin layer of about $0.2 d$~\cite{BasuZhang2009,BasuZhang2011}. That means, that when constructing for example a near-field thermophotovoltaic device choosing $d = 100\,{\rm nm}$ most energy is already absorbed in a thin surface layer of about $20\,{\rm nm}$. This is very unfavorable for applications in near-field thermophotovoltaic devices, since only the electron-hole pairs in this thin layer can effectively be used for energy conversion~\cite{ParkEtAl2007}.

As could be shown recently~\cite{BiehsEtAl2012,GuoEtAl2012} for so called hyperbolic or indefinite materials~\cite{SmithSchurig2003}, which exist naturally~\cite{DrachevEtAl2013,ThompsonEtAl1998,CaldwellEtAl2014} but can also be constructed artificially by combining layers of a dielectric and a plasmonic/polaritonic material~\cite{HoffmanEtAl2007,KrishnamoorthyEtAl2012,LangEtAl2013} or by using plasmonic/polaritonic nanowire structures~\cite{SmithSchurig2003,WangbergEtAl2006,YaoEtAl2008,NoginovEtAl2009,CaiEtAl2010,DrachevEtAl2013,PoddubnyEtAl2013}, the nanoscale heat radiation by hyperbolic modes can result in heat fluxes which are on the order of or even larger than the heat flux by surface modes~\cite{BiehsEtAl2012}. This is due to a broad band of hyperbolic modes which are, in fact, frustrated total internal reflection modes. Recently, hyperbolic structures were proposed for applications in nTPV~\cite{Nefedov2011,Simovski2013}. 

Having a broad frequency band for nanoscale heat radiation seems to be disadvantageous for nTPV, but this disadvantage is compensated by a striking property of hyperbolic modes: hyperbolic modes are propagating modes inside the hyperbolic metamaterials and therefore it can be expected that the penetration depth is much larger than for surface modes. Hence, the effective layer on which electron-hole pairs are generated can be orders of magnitude larger than for surface-mode driven heat transfer. This property could be shown by means of an effective description, presented very recently by us~\cite{Slawa2014}. But such effective descriptions should be taken with care for describing near-field thermal radiation, since it tends to overestimate the hyperbolic contribution to the heat flux~\cite{LiuShen2013,BiehsEtAl2013} and it does not correctly describe the surface modes of the composite materials of the hyperbolic structure~\cite{BiehsEtAl2013,TschikinEtAl2013}.

In this paper, we study the penetration depth of the energy flow in multilayer hyperbolic materials using an exact S-matrix method~\cite{AuslenderHava1996,Francoeur2009} based on the Green's function~\cite{Sipe} formalism combined with fluctuational electrodynamics~\cite{RytovBook1989}. We have previously shown that the attenuation length can be very large for hyperbolic nanowire and multilayer materials using an effective medium description~\cite{Slawa2014}. Here, we use the exact formalism to study the energy flux and penetration depth for multilayer hyperbolic metamaterials. The exact formalism is compared with effective medium theory to better understand potential limitations of the latter. We emphasize that the here developed method can directly be used to make exact calculation for the energy streamlines inside hyperbolic multilayer structures which were treated only within the effective medium approach so far~\cite{BrightEtAl2014}.
%Here, by using an exact formalism for multilayer structures, only, we prove that the penetration depth inside hyperbolic multilayer materials is indeed much larger than in materials where the heat flux is dominated by surface modes. It can be even larger than predicted by the effective medium theory. Furthermore, our formalism allows for calculating the energy flux inside any kind of multilayer structure due to the radiative heat transfer (in far and near field) and therefore it enables one to determine the volume of the absorber in which the thermal radiation is dissipated. Further, it allows for studying deviations from the effective medium theory systematically.

%The paper is organized as follows: In Sec. 2 we briefly outline fluctuational electrodynamics for determining the radiative heat transfer expression. In Sec. 3 we define the spectral and total attenuation length and introduce the considered geometries together with the effective medium theory. We discuss the damping inside a multilayer hyperbolic material and compare numerical results of our exact formalism for a GaN/Ge multilayer structure with the predictions of the effective medium theory. Finally, we give a brief conclusion in Sec. 4.

%%%%%%%%%%%%%%%%%%%%%%%%%%%%%%%%%%%%%%%%%%%%%%%%%%%%%%%%%%%%
\section{Energy flux inside a layered medium}
%%%%%%%%%%%%%%%%%%%%%%%%%%%%%%%%%%%%%%%%%%%%%%%%%%%%%%%%%%%%

The theoretical description of near-field heat radiation is in most studies based on fluctuational electrodynamics~\cite{RytovBook1989}. Within this theory it is assumed that the thermal fluctuating fields of a dielectric body which is assumed to be in local thermal equilibrium at a temperature $T$ are on a macroscopic scale due to fluctuational source current densities. Hence, Maxwell's equations are augmented by fluctuational Gaussian source currents $\mathbf{J}^{\rm m}$ and $\mathbf{J}^{\rm e}$ yielding
\begin{align}
	\nabla\times\mathbf{E}(\mathbf{r},t) &= -\mathbf{J}^{\rm m}(\mathbf{r},t) -\frac{\partial \mathbf{B}(\mathbf{r},t)}{\partial t}, \\
  \nabla\times\mathbf{H}(\mathbf{r},t) &= \mathbf{J}^{\rm e}(\mathbf{r},t) + \frac{\partial \mathbf{D}(\mathbf{r},t)}{\partial t}.
\end{align}
For nonmagnetic materials the fluctuating magnetic source currents can be neglected $\mathbf{J}^{\rm m}(\mathbf{r},t) = \mathbf{0}$. The source current density $\mathbf{J}^{\rm e}(\mathbf{r},t)$ is assumed to have zero mean value $\langle \mathbf{J}^{\rm e} \rangle = \mathbf{0}$, where the brackets symbolize the ensemble average. Then it is further assumed that the second moment or correlation function of the source currents is given by the fluctuation dissipation theorem of second kind~\cite{Kubo}
\begin{equation}
	\langle J_\alpha^{\rm e} (\mathbf{r},\omega) J_\beta^{\rm e} (\mathbf{r}',\omega') \rangle = 4 \pi \omega  \Theta(\omega,T) \epsilon_{\rm vac} \epsilon_{\alpha\beta}'' \delta(\mathbf{r - r'}) \delta (\omega + \omega'),
\end{equation}
where $\Theta(\omega,T) = \hslash \omega /({\rm e}^{\frac{\hslash \omega}{k_{\rm B} T}} -1)$ and $\epsilon_{\alpha\beta}''$ is the imaginary part of the permittivity tensor of the considered material; $\epsilon_{\rm vac}$ is the permittivity of vacuum, $2 \pi \hslash$ is Planck's constant, $\kb$ is Boltzmann's constant, $\omega$ is the circular frequency, and $\delta$ stands for the delta function. Here, obviously quantum mechanics in form of the fluctuation dissipation theorem enters into the theoretical description which can therefore be regarded as a semi-classical theory. However, a full quantum mechanical description agrees with this method~\cite{JanowiczEtAl2003}.

Now, since the fields are linearly related to the sources they can be expressed as
\begin{align}
	\mathbf{E}(\mathbf{r},\omega) &= \ri \omega \mu_{\rm vac} \int_V \!\! \rd^3 r'\, 
            \GE(\mathbf{r,r'};\omega) \cdot \mathbf{J}^{\rm e}(\mathbf{r}',\omega), \\
	\mathbf{H}(\mathbf{r},\omega) &= \ri \omega \mu_{\rm vac} \int_V \!\! \rd^3 r'\, 
          \GH(\mathbf{r,r'};\omega) \cdot \mathbf{J}^{\rm e}(\mathbf{r}',\omega)
\end{align}
introducing the dyadic Green's functions $\GE$ and $\GH$. Since we only consider nonmagnetic materials $\mu_{\rm vac}$ is the permeability of vacuum and of all materials. 
By means of the fluctuation dissipation theorem we can now derive the mean Poynting vector or Maxwell's stress tensor, for instance. For some general elaborations on the stress tensor and the Poynting vector within the formalism of fluctuational electrodynamics we refer the interested reader to ~\cite{Arvind2014}. Since we are interested in heat radiation we focus on the Poynting vector.

\begin{figure}[htb]
	\centering\includegraphics[width=8cm]{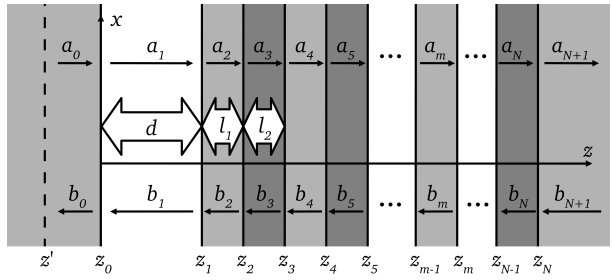}
	\caption{Sketch of the considered geometry: for $z<z_0$ halfspace filled with GaN; for $z_0<z<z_1$ vacuum gap of width $d$; for $z_1<z<z_N$ bilayer structure with a periode $\Lambda=l_1 + l_2$. The width of the GaN (Ge) layers is $l_1$ ($l_2$). For $z>z_N$ halfspace filled with GaN.}
	\label{Fig:LayeredMedium}
\end{figure}

Let us now assume that we have a situation as depicted in Fig.~\ref{Fig:LayeredMedium}. For $z < z_0 = 0$ we have a semi-infinite isotropic material which is at local thermal equilibrium at temperature $T_0$. This halfspace is separated by a vacuum gap of size $d$ from a second halfspace which can be any kind of multilayer structure and  which is assumed, for sake of clarity, to be at zero temperature. This assumption means that this medium does not emit thermal photons but it can only scatter and absorb them.  However, of course, this medium could be set at any temperature. Straight forwardly we obtain the expression (using the Einstein convention)
\begin{equation}
	\langle S_z \rangle =  2 \Re \int\limits_{0}^{\infty} \mathrm d\omega \,\Theta(\omega, T_0) \frac{\mu_{\rm vac}^2\omega^3 \operatorname{Im}(\epsilon_{0})}{\pi} 
    \int\limits_{z'<0} \mathrm d^3\bf {r'} \epsilon_{z\alpha\beta}(\mathbb{G}^{\rm E}({\bf r}, {\bf r'})\cdot {\mathbb{G}^{\rm H}}^\dagger ({\bf r}, {\bf r'}))_{\alpha\beta},  
	\label{Eq:PoyntingVectorGeneral}
\end{equation} 
where $ \epsilon_{z\alpha\beta}$ is the antisymmetric Levi-Civita tensor and $\mathds{G}^{\rm E}({\bf r}, {\bf r'})$ ($\mathds{G}^{\rm H}({\bf r}, {\bf r'})$) are the electric (magnetic) dyadic Green's functions of the considered geometry with source points $\mathbf{r}'$ in the halfspace for $z < 0$ and the observation point $\mathbf{r}$ inside the vacuum gap or the second halfspace for $z > d$. Hence, when knowing the dyadic Green's functions we can determine the Poynting vector which describes the energy transfer by thermal emission at any position within the multilayer structure ($z > d$) and for any separation distance $d$. Note, that $\epsilon_0$ is the permittivity inside the halfspace for $z < 0$.

To determine the mean Poynting vector describing the energy flow, we need to determine the corresponding dyadic Green's functions for the structure depicted in Fig.~\ref{Fig:LayeredMedium}. These dyadic Green's function can be determined by a standard procedure~\cite{Sipe}. When inserting these expressions into~(\ref{Eq:PoyntingVectorGeneral}) we obtain
\begin{eqnarray}
	\hspace{-0.5 cm}\langle S_z \rangle =  \int\limits_{0}^{\infty} \frac{\mathrm d\omega}{2\pi} \Theta(\omega, T_0) \sum_{j={\rm s,p}} \int\limits_{0}^{\infty}  \frac{\mathrm d ^2\kappa}{(2\pi)^2}  \mathfrak{T}^{(j)}(\omega,\kappa;z),
	\label{EQ:Poynting}
\end{eqnarray}
where we have introduced the transmission coefficient in polarization $j$ ($j$=s,p) of each mode $(\omega, \kappa)$ at a distance $z$ from the surface as
\begin{equation}
	\begin{split}
		\mathfrak{T}^{({j})}(\omega,\kappa;z) &= \frac{ \gamma_{0}^{'}}{|\gamma_{0}|^2} \biggl[ \mathrm{Re}(c_{n}^{(j)}) \Bigl( {\rm e}^{-2\gamma_n^{''}z} |a_{n}^{(j)}|^2 - {\rm e}^{2\gamma_n^{''}z} |b_{n}^{(j)}|^2 \Bigr) \\
           &\qquad+  {\rm i}\, \mathrm{Im}(c_{n}^{(j)})\Bigl( {\rm e}^{2{\rm i}\gamma_n^{'}z} a_{n}^{(j)} {b_{n}^{(j)}}^{*} - {\rm e}^{-2{\rm i}\gamma_n^{'}z} {a_{n}^{(j)}}^{*} b_{n}^{(j)}   \Bigr)  \biggr]  
	\end{split}
	\label{EQ:TC}
\end{equation}
for $z_{n-1}<z<z_n$. The coefficients in Eq.~(\ref{EQ:TC}) are 
\begin{equation}
	c_n^{(\rm s)} = \gamma_n, \qquad\text{and} \qquad
	c_n^{(\rm p)} = \frac{\kappa^2 +|\gamma_0|^2}{|k_0|^2}\frac{\gamma_n{k_n^*}^2}{|k_n|^2}
\end{equation} 
with $k_n^2=\epsilon_n\frac{\omega^2}{c^2}=\kappa^2 +\gamma_n^2$ the square of the wave vector and $\epsilon_n$ the permittivity in the $n$-th layer; $c$ is the vacuum speed of light, $\kappa=\sqrt{k_x^2 +k_y^2}$ the wave vector component parallel to the surface and $\gamma_n=k_{z,n}$ the wave vector component in $z$ direction in the $n$-th layer. $'$ and $''$ donate the real and imaginary part and $*$ the complex conjugate of a number. The amplitudes $a_n^{(j)}$ and $b_{n}^{(j)}$ are determined by the S-matrix method described in Appendix~\ref{AppendixSmatrix}. In the same manner as detailed above, one can determine the heat flux from the medium at $z > d$ which is assumed to have a temperature $T_2$. In this case we obtain the same result as in Eq.~(\ref{EQ:Poynting}) but with $\Theta(\omega, T_0)$ being replaced by $-\Theta(\omega, T_2)$. Then the total heat flux is the sum of both contributions.

%%%%%%%%%%%%%%%%%%%%%%%%%%%%%%%%%%%%%%%%%%%%%%%%%%%%%%%%%%%%
\section{Attenuation length of heat flux}
%%%%%%%%%%%%%%%%%%%%%%%%%%%%%%%%%%%%%%%%%%%%%%%%%%%%%%%%%%%%

Let us assume that the temperature of the first halfspace ($z < 0$) is $T_0 = T+ \Delta T$ and that of the multilayer structure ($z > d$) is $T_2 = T$ with $\Delta T \ll T$. One can determine the heat transfer coefficient from~(\ref{EQ:Poynting}) 
\begin{equation}
	h(z) =  \int\limits_{0}^{\infty} \frac{\mathrm d\omega}{2\pi} f(\omega,T) \sum_{j={\rm s,p}}\overline{\mathfrak{T}}^{(j)}(\omega,z) 
        = \int\limits_{0}^{\infty} \frac{\mathrm d\omega}{2\pi} H(\omega,z) 
\end{equation}
where $f(\omega,T)=(\hslash \omega)^2/(k_{\rm B}T^2) {\rm e}^{\hslash \omega/k_{\rm B}T}/({\rm e}^{\hslash \omega/k_{\rm B}T}-1)^2$ and $H(\omega,z)$ is the spectral heat transfer coefficient. In this equation $\overline{\mathfrak{T}}^{(j)}(\omega,z)=\int\limits_{0}^{\infty}  \frac{\mathrm d ^2\kappa}{(2\pi)^2}  \mathfrak{T}^{(j)}(\omega,\kappa;z)$ is the mean transmission coefficient of all modes at the frequency $\omega$ over the distance $z$. The energy flux is then given by $h \Delta T$. 
By means of this expression we can define the spectral attenuation length $L_{\rm a}$ as the distance $z$ inside the multilayer structure ($z > d$) at which the spectral heat transfer coefficient $H(\omega,z)$ has dropped to $H(\omega,d)$/e. The total attenuation length  $l_{\rm a}$ is similarly defined as the distance $z$inside the multilayer structure ($z > d$) at which the heat transfer coefficient $h(z)$ has dropped to $h(d)$/e. It is clear from its definition that the asymptotic behavior of heat transfer coefficient  at long distance is exponentially decaying. However as we are going to see in the next section, compared with bulk homogeneous materials, the attenuation length of heat flux can be significantly increased in particular layered structures.

%%%%%%%%%%%%%%%%%%%%%%%%%%%%%%%%%%%%%%%%%%%%%%%%%%%%%%%%%%%%
\subsection{Heat flux damping inside a layered hyperbolic medium}
%%%%%%%%%%%%%%%%%%%%%%%%%%%%%%%%%%%%%%%%%%%%%%%%%%%%%%%%%%%%

\begin{figure}[htb]
	\centering\includegraphics[width=8cm]{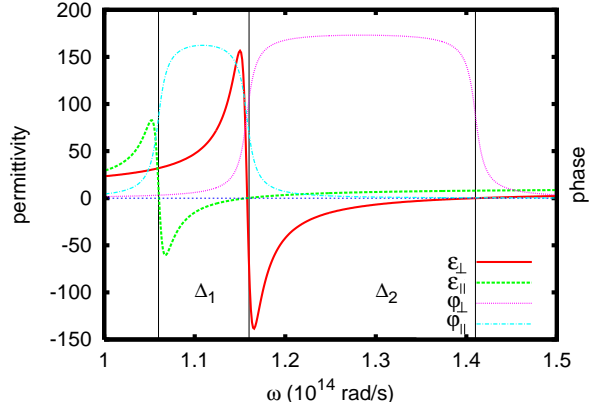}
	\caption{Plot of the real parts of the permittivities $\epsilon_\parallel$ and $\epsilon_\perp$ and their phases $\varphi_\parallel$
and $\varphi_\perp$ from Eqs.~(\ref{Eq:epsilonparallel}) and (\ref{Eq:epsilonperp}) in $^\circ$ for an effective GaN/Ge multilayer structure 
choosing $f = 0.5$. The vertical lines mark the edges of the two hyperbolic bands $\Delta_1$ and $\Delta_2$.\label{Fig:permittivity}}
	\label{Fig:Epsilon}
\end{figure}

The formalism introduced above is general and it could be applied to describe heat transport by radiation through any arbitrary layered structures. We focus here our attention on  specific media called hyperbolic media. Those media support modes that are governed by an hyperbolic dispersion relation. In this paper we consider hyperbolic media composed of alternated layers of materials (see  Fig.~\ref{Fig:LayeredMedium}) whose real parts of dielectric permittivities $\epsilon_1$ and $\epsilon_2$ are of opposite sign in a given spectral range~\cite{SmithSchurig2003}.
According to the effective medium theory, in the longwavelength approximation, the structure is analog to an uniaxial crystal~\cite{Yeh} with a permittivity tensor $\boldsymbol{\epsilon}=\epsilon_{\parallel}(\mathbf{e}_x\otimes\mathbf{e}_x+\mathbf{e}_y\otimes\mathbf{e}_y)+\epsilon_{\perp}\mathbf{e}_z \otimes \mathbf{e}_z$ of component 
\begin{align}
  \epsilon_\parallel  &= f\epsilon_{\rm 1} + (1-f)\epsilon_{\rm 2}. \label{Eq:epsilonparallel}
\end{align}
in the direction parallel to the surface and
\begin{align}
  \epsilon_\perp &= \frac{\epsilon_{\rm 1}\epsilon_{\rm 2}}{f\epsilon_{\rm 2} + (1-f)\epsilon_{\rm 1}}, \label{Eq:epsilonperp}
\end{align}
along the optical axis $\mathbf{e}_z$. $f$ is the volume filling factor of medium 1 which is in our case Germanium (Ge). In Fig.~\ref{Fig:Epsilon} these components are plotted in the case of a Gallium Nitride/Germanium (GaN/Ge) multilayer structure. It can be seen that there are two frequency bands named $\Delta_1$ ($1.06\cdot10^{14}\,{\rm rad/s} < \omega < 1.16\cdot10^{14}\,{\rm rad/s}$) and $\Delta_2$ ($1.16\cdot10^{14}\,{\rm rad/s} < \omega < 1.41\cdot10^{14}\,{\rm rad/s}$) where the product $\Re(\epsilon_\parallel) \Re(\epsilon_\perp) < 0$. 
These correspond to the hyperbolic frequency bands which allow for propagating p-polarized waves with hyperbolic isofrequency curves instead of elliptical ones.
The $z$ components of the effective wave vectors for s- and p-polarized waves are solutions of the vector wave equation~\cite{Yeh} and are given by 
\begin{align}
	\gamma_\rs&=\sqrt{\omega^2/c^2 \epsilon_{\parallel}-\kappa^2}, \label{Eq:gammas} \\ 
	\gamma_\rp&=\sqrt{\omega^2/c^2 \epsilon_{\parallel}-\kappa^2\epsilon_{\parallel}/\epsilon_{\perp}} \label{Eq:gammap}.
\end{align}

It is worth noting that the attenuation length of heat flux through a homogenized structure is naturally related to the penetration depth of the intensity $\delta_j=\frac{1}{2 \operatorname{Im}(\gamma_j)}$ $(j=\rs,\rp)$ of electric and magnetic fields. To get some insight on the flux attenuation mechanism in hyperbolic media we examine below how a plane wave traveling along the $z$ direction is damped. Since these hyperbolic modes are p polarized only, we focus on the damping of p polarized waves.

Using the polar representation $\epsilon_{\parallel}=|\epsilon_{\parallel}|e^{\ri{\varphi_{\parallel}}}$ and $ \epsilon_{\perp}=|\epsilon_{\perp}|e^{\ri{\varphi_{\perp}}}$ the $z$ component of the wave vector in polarization p can be recast as
\begin{eqnarray}
 \gamma_\rp&=|\epsilon_{\parallel}|^{1/2}e^{\ri\frac{\varphi_{\parallel}}{2}}k_1&\sqrt{1-\frac{\kappa^2}{k_1^2}\frac{e^{- \ri \varphi_{\perp}}}{|\epsilon_{\perp}|}},
\end{eqnarray}
where $k_1=\frac{\omega}{c}$. From this expression we obtain for $\kappa \gg k_1 \sqrt{|\epsilon_\perp|}$
\begin{eqnarray}
  \gamma_\rp &\approx \ri\kappa\sqrt{\frac{|\epsilon_\parallel|}{|\epsilon_{\perp}|}}e^{\ri ({\varphi_{\parallel}}-{\varphi_{\perp}})/2}.
\end{eqnarray}
Taking the imaginary part of this expression yields 
\begin{equation}
  \Im(\gamma_\rp) \approx \kappa \sqrt{\frac{|\epsilon_\parallel|}{|\epsilon_{\perp}|}} \cos\biggl(\frac{\varphi_\parallel-\varphi_\perp}{2}\biggr)
\end{equation}
which determines the damping of a plane wave travelling in $z$ direction inside the uni-axial material. Therefore we have small damping inside the anisotropic material (compared to the isotropic case, where $\epsilon_\parallel = \epsilon_\perp$ and hence $\Im(\gamma_\rp) \approx \kappa$) if 
\begin{equation}
  \varphi_\parallel -\varphi_\perp = \pm \pi
\end{equation}
or if $|\epsilon_\parallel| \ll |\epsilon_{\perp}|$. In particular inside a hyperbolic material, where $\Re(\epsilon_\parallel) \Re(\epsilon_\perp) < 0$ the condition on the phases can be fullfilled if the imaginary parts of the permittivities perpendicular and parallel to the optical axis are small, i.e.\ if $\epsilon_\perp'' / |\epsilon_\perp'| \ll 1$ and $\epsilon_\parallel'' / \epsilon_\parallel' \ll 1$. On the other hand, there is also small damping for strong anisotropic materials with $|\epsilon_\parallel| \ll |\epsilon_{\perp}|$. 

In the opposite limit where $\kappa \ll k_1 \sqrt{|\epsilon_\perp|}$ we find
\begin{equation}
  \gamma_\rp \approx |\epsilon_\parallel|^{1/2}e^{\ri\frac{\varphi_\parallel}{2}}k_1\,\biggl(1-\frac{\kappa^2}{2 k_0^2}\frac{e^{-\ri \varphi_{\perp}}}{|\epsilon_{\perp}|}\biggr).
\end{equation}
and therefore
\begin{equation}
  \Im(\gamma_\rp) \approx 
        |\epsilon_\parallel |^{1/2} k_1 \sin\bigl({\frac{\varphi_\parallel}{2}}\bigr) . 
\end{equation}
Hence, if $\varphi_\parallel \approx 0$ or $|\epsilon_\parallel| \approx 0$, i.e. losses parallel to the interface are small, we have a large penetration of fields. Note that if $|\epsilon_\parallel|$ is small the damping is in both limits small as well.

\begin{figure}[htb]
	\begin{center}
		\includegraphics[width=0.45\textwidth]{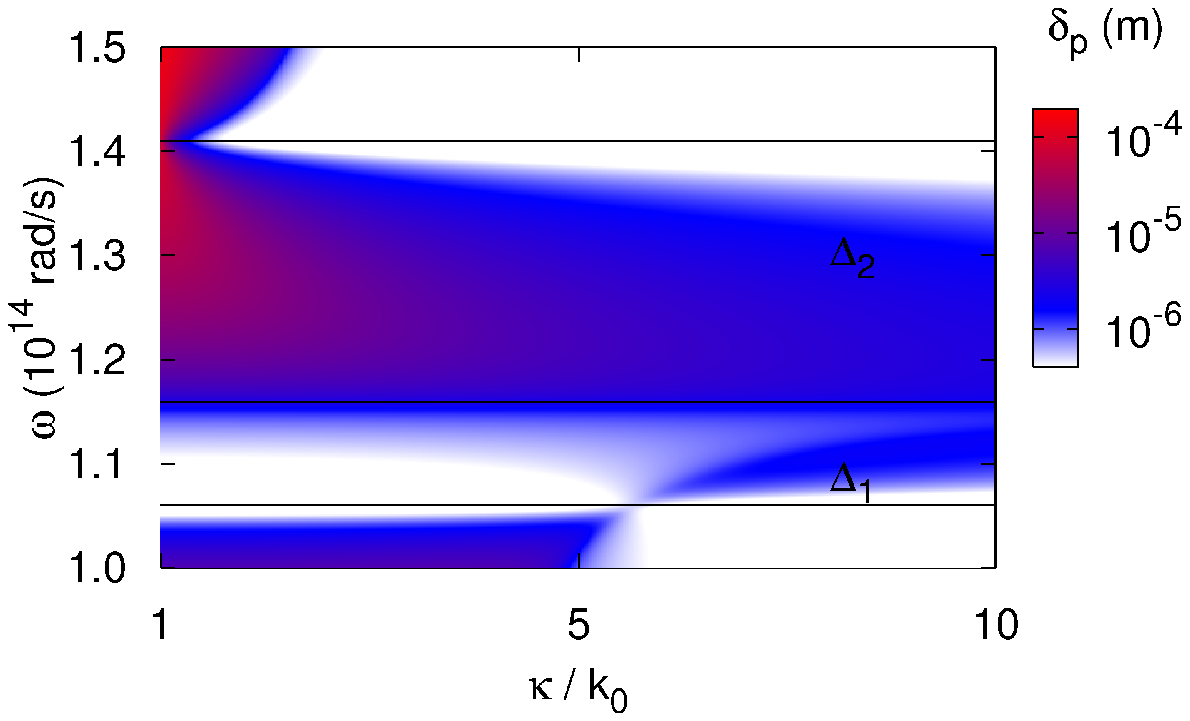}
		\includegraphics[width=0.45\textwidth]{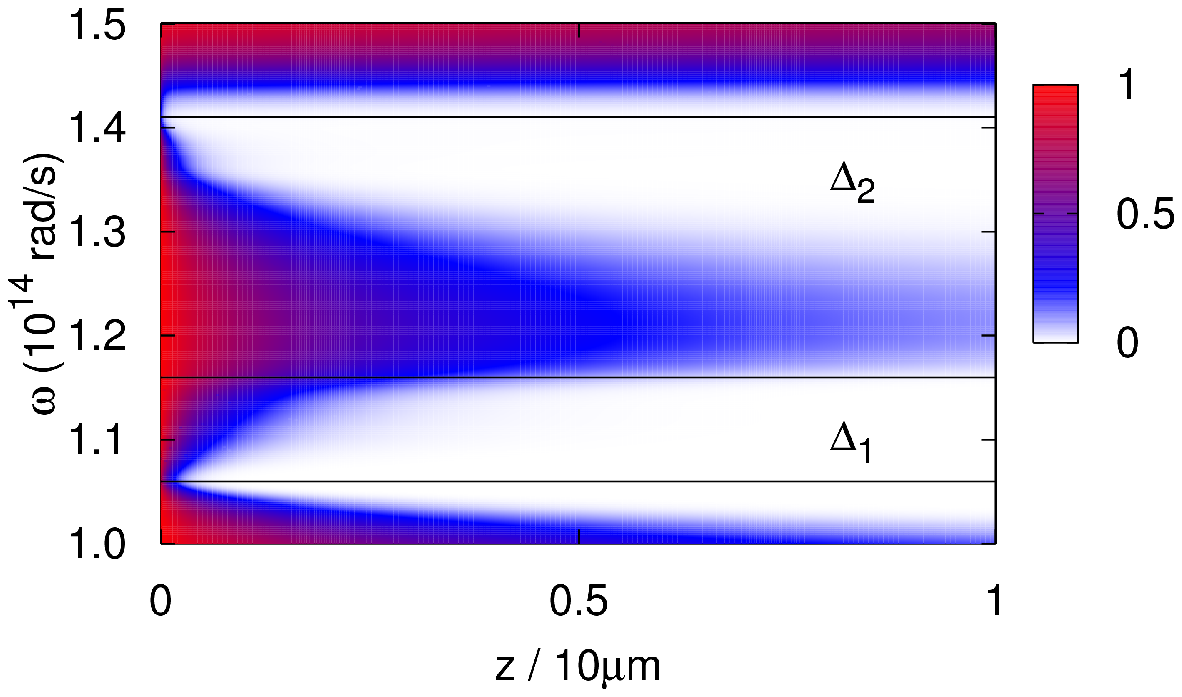}
	\end{center}
	\caption{Plot of $\delta_{\rm p} = 1/ (2 \Im(\gamma_\rp))$ (left) as a function of frequency and $\kappa$ and the normalized spectral 
heat transfer coefficient $H(\omega,z)/H(\omega,z = d)$ (right) with respect to the frequency $\omega$ and distance $z$ inside the hyperbolic material. The same
parameters as in Fig.~\ref{Fig:Epsilon} are used. The temperature is $T = 300\,{\rm K}$ and $d = 100\,{\rm nm}$.}
	\label{Fig:ImagKz}
\end{figure}

%%%%%%%%%%%%%%%%%%%%%%%%%%%%%%%%%%%%%%%%%%%%%%%%%%%%%%%%%%%%
\subsection{Results and discussion - finite doublelayer structure}
%%%%%%%%%%%%%%%%%%%%%%%%%%%%%%%%%%%%%%%%%%%%%%%%%%%%%%%%%%%%

Now, we discuss these mechanisms in the infrared range for a simple GaN/Ge periodic structure. For having the broadest possible hyperbolic bands and therefore the largest hyperbolic effect we consider identical widths for the two unit layers, i.e.\ the filling factor is $0.5$. To evaluate the thermal performances of this medium, we assume that its left side is located at a distance $z=d$ from a bulk GaN halfspace maintained at temperature $T = 300\,{\rm K}$ and we calculate the heat transfer coefficients through the layered structure  for different separation gap. 
The finite multilayer material is assumed to be on a semi-infinite substrate ($z>z_N$) made of the same material as the left halfspace. In the frequency range of interest, the permittivity of Ge layers is set to $\epsilon_{\rm Ge}=16$ while for the polar material Gallium Nitride (GaN) it is well described by the Drude-Lorentz model~\cite{Adachi}
\begin{equation}
	\epsilon_{\rm GaN}(\omega)=\epsilon_{\infty}\frac{\omega_{\rm LO}^2 - \omega^2 -{\rm i}\gamma\omega}{\omega_{\rm TO}^2 - \omega^2 -{\rm i}\gamma\omega},
\end{equation}
where the permittivity at infinite frequency, the damping coefficient, the transverse and longitudinal optical phonon frequencies are given by $\epsilon_{\infty}=5.35$, $\gamma=1.52\cdot10^{12}\,$rad/s, $\omega_{\rm TO}=1.06\cdot10^{14}\,$rad/s and $\omega_{\rm LO}=1.41\cdot10^{14}\,$rad/s, respectively. In Fig.~\ref{Fig:ImagKz} the inverse damping factor $\delta_\rp = 1/ (2 \Im(\gamma_\rp))$ as a function of $\omega$ and $\kappa$ as well as the spectral heat transfer coefficient $H(\omega,z)$ as a function of frequency $\omega$ and distance $z$ inside the metamaterial are plotted. It becomes obvious that the damping factor inside the hyperbolic bands and especially inside the band $\Delta_2$ is very small because of the above discussed fact that the penetration depth is large in the hyperbolic bands where $|\epsilon_\parallel| \ll |\epsilon_\perp|$. In particular, the penetration depth of the heat flux inside the material is slightly smaller than $10$ microns, i.e.\ comparable to the thermal wavelength $\lambda_{\rm th} = \hslash c / \kb T$ which is about $7.6\,\mu {\rm m}$ at $T = 300\,{\rm K}$.

\begin{figure}[htb]
  \centering\includegraphics[width=0.45\textwidth]{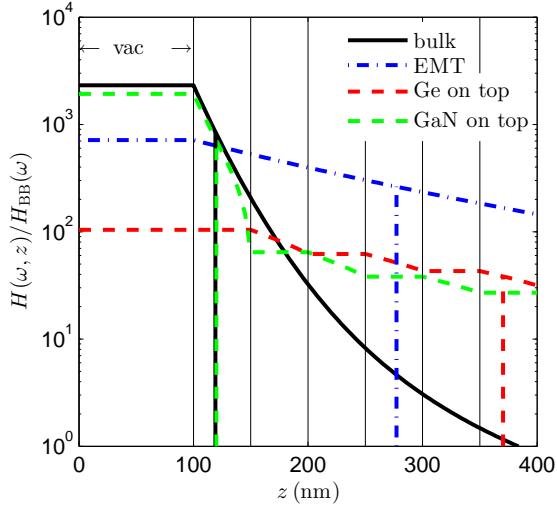}
  \caption{Spectral heat transfer coefficient $H(\omega,z)$ for the surface mode resonance frequency of GaN $\omega=1.36\cdot 10^{14}\,$rad/s, $d=100\,$nm and $\Lambda=100\,$nm versus the distance $z$. The spectral heat transfer coefficient $H(\omega,z)$ is normalized to the black body value $H_{\rm BB}(\omega) = f(\omega,T) \omega^2/(2 \pi c^2)$. The dashed vertical lines mark the distance where $H(\omega,z) = H(\omega,d)/$e for each case. The thin solid vertical lines are the interfaces of the multilayers. Here and in the following we use $N = 40$, i.e.\ we have $20$ bilayers.}
  \label{Fig:sHTCinMed}
\end{figure}

To compare these results with the exact calculations based on the scattering matrix theory we have plotted in Fig.~\ref{Fig:sHTCinMed} the spectral heat transfer coefficient $H(\omega,z)$ in the gap and in the material ($z > d = 100\,{\rm nm}$) inside the hyperbolic band $\Delta_2$ for the following configurations: First, the halfspace at $z < 0$ is given by a bulk GaN halfspace. For the halfspace at $z > d$ we consider (i) bulk GaN, (ii) an effective infinite GaN/Ge doublelayer material (using the EMT for calculations), (iii) a finite GaN/Ge doublelayer material with Ge as topmost layer, and (iv) a finite GaN/Ge doublelayer material with GaN as topmost layer. For (iii) and (iv) the finite doublelayer structure is on top of a GaN half space.
For numerical reasons and the fact that real structures are finite we consider $N = 40$ layers for the exact numerical calculations. The black solid line shows the exponential decay of $H(\omega,z)$ for increasing $z$ in the bulk case (i) where the heat flux is dominated by the surface modes (the surface phonon-polariton supported by the GaN sample). The blue dot-dashed line shows the heat transfer coefficient inside a homogenized medium (ii). The green and red dashed lines are the exact results for the  layered GaN/Ge medium with Ge (iii) or GaN (iv) as topmost layer. In both cases the heat transfer coefficient is constant inside the Ge layer due to the negligible dissipation inside Ge. The solid vertical lines represent the interfaces of the multilayers and the vertical dashed lines represent the distance at which the spectral heat transfer has dropped to 1/e of its value at the interface, i.e.\ it marks the attenuation length $L_\ra$ at the given frequency and distance. It can be seen that in the case of bulk GaN (i) the spectral heat transfer coefficient at the surface (at $z=d$) is larger than in the cases of the layered hyperbolic metamaterial structure (ii)-(iv). On the other hand the attenuation length $L_{\rm a}$ is much smaller for (i) compared to (ii) and (iii). Note that the multilayer structure (iv) with GaN as topmost layer has almost the same properties (regarding the exchanged heat flux as well as the attenuation length) as bulk GaN~\cite{Volokitin2001,Biehs2007,MillerEtAl2014}.

\begin{figure*}[htb]
  \begin{center}
		\includegraphics[width=0.45\textwidth]{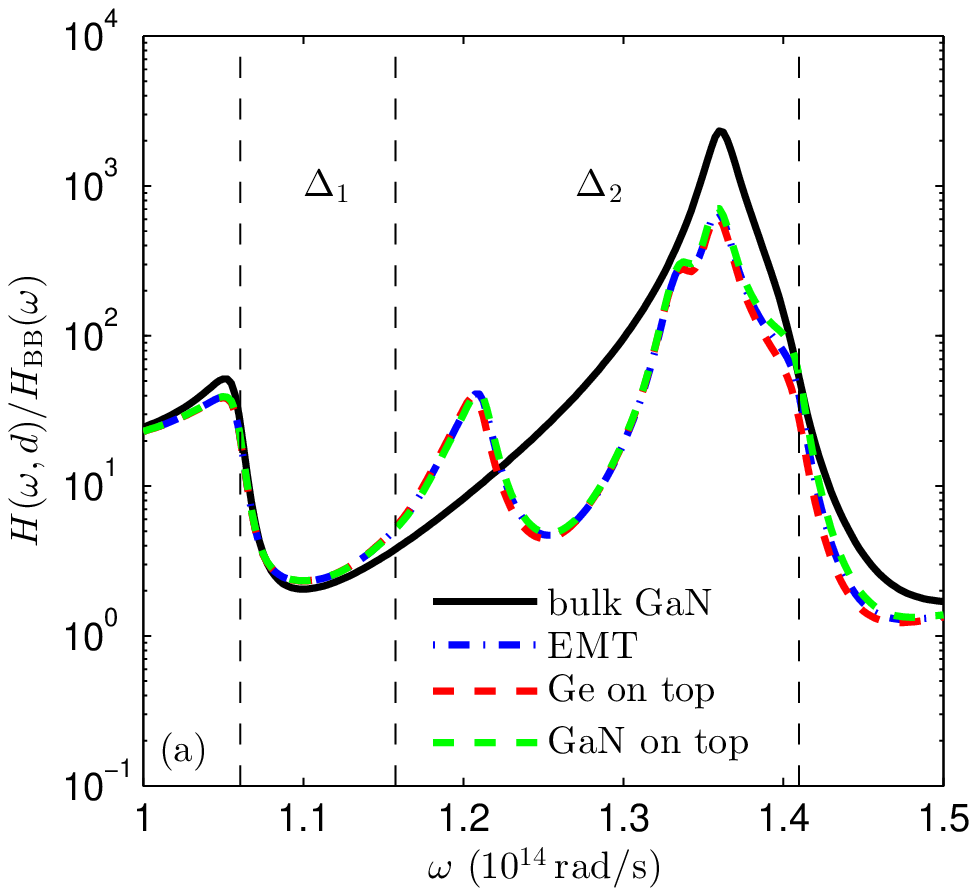}	\includegraphics[width=0.45\textwidth]{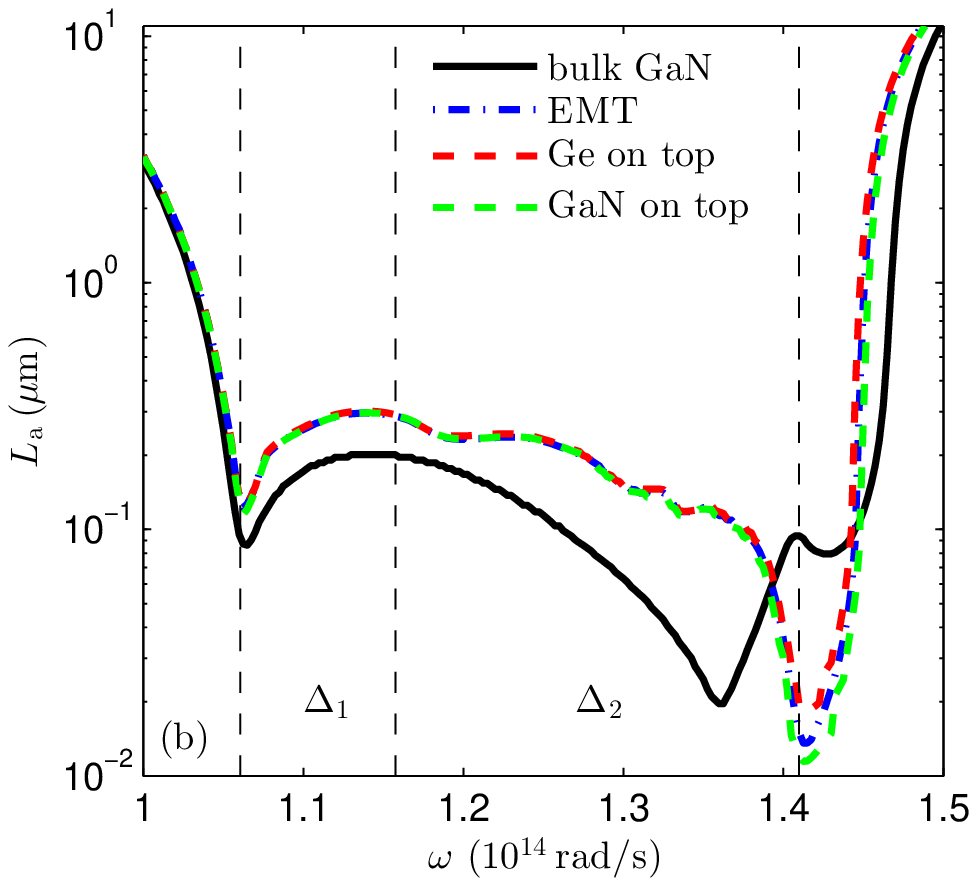} 
		\includegraphics[width=0.45\textwidth]{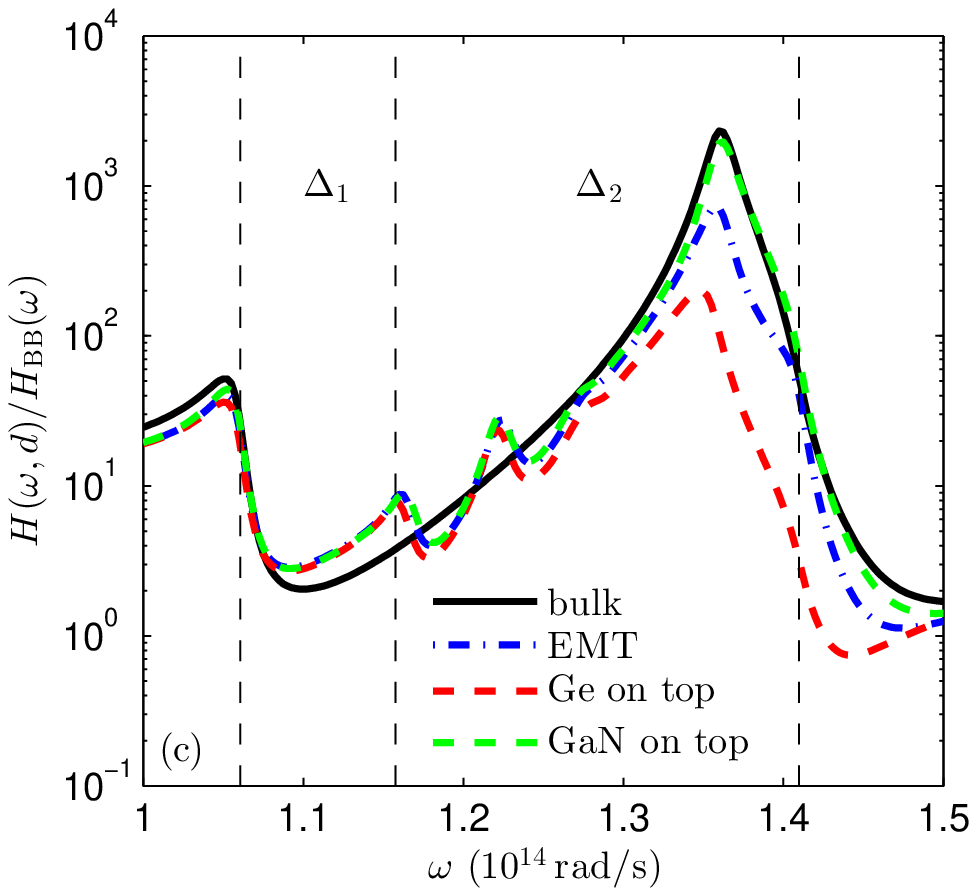}	\includegraphics[width=0.45\textwidth]{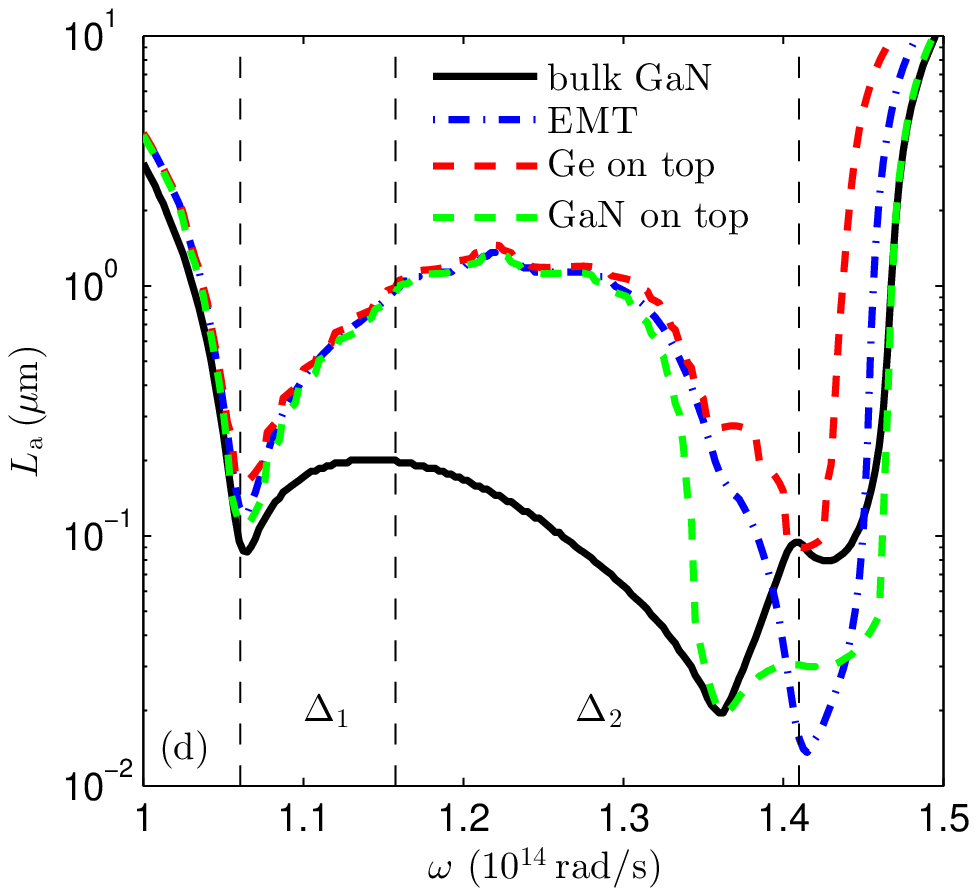} 
  \end{center}
  \caption{ (a) and (c) show the spectral heat transfer coefficient $H(\omega,d)$ normalized to the black body value $H_{\rm BB}(\omega) = f(\omega,T) \omega^2/(2 \pi c^2)$ for $\Lambda = 10\,{\rm nm}$ (top) and $\Lambda = 100\,{\rm nm}$ (bottom) keeping the distance fixed at $d=100\,$nm. (b) and (d) show the spectral attenuation length $L_\ra$.}
	\label{Fig:sHTC}
\end{figure*}

The frequency dependence of heat transfer coefficient and of attenuation lengths is described in Fig.~\ref{Fig:sHTC}  for the four cases (i)-(iv) for a separation gap  $d = 100\,{\rm nm}$ which corresponds to a distance where heat transfer occurs mainly due to near-field interaction. Note, that the attenuation length for bulk GaN is inside the {\itshape reststrahlen} band ($\omega_{\rm TO} < \omega < \omega_{\rm LO}$) smaller than $200\,{\rm nm}$ which is due to the strong damping of the surface modes. It can be seen that inside the {\itshape reststrahlen} band where hyperbolic modes and surface modes exist, the hyperbolic structures have an attenuation length which is up to one order of magnitude larger than for bulk GaN. Further, the attenuation length inside the hyperbolic bands scales with the size of the hyperbolic structure which is $200\,{\rm nm}$ for $\Lambda = 10\,{\rm nm}$ and $2\,\mu{\rm m}$ for  $\Lambda = 100\,{\rm nm}$. For larger structures one can expect to have an even larger attenuation length as indicated by the effective medium result in Fig.~\ref{Fig:ImagKz}. Indeed, in this case an attenuation length on the order of three microns within the hyperbolic region can be found~\cite{Slawa2014}.

On the other hand the spectral heat transfer coefficient is for all structures very similar but for bulk GaN the peak at the surface mode frequency is more pronounced than for the hyperbolic structures. Finally, the deviation between the exact and effective results is small for $\Lambda = 10\,{\rm nm}$ as can be expected, since $\Lambda \ll d$ in this case. However, for $\Lambda = 100\,{\rm nm}$, i.e.\ $\Lambda \approx d$ the deviations between the effective and exact description become important. In particular, the choice of the material of the first layer has a large impact as discussed in detail in ~\cite{BiehsEtAl2013,TschikinEtAl2013,KidwaiEtAl2012}. 

\begin{figure*}[htb]
  \begin{center}
		\includegraphics[width=0.45\textwidth]{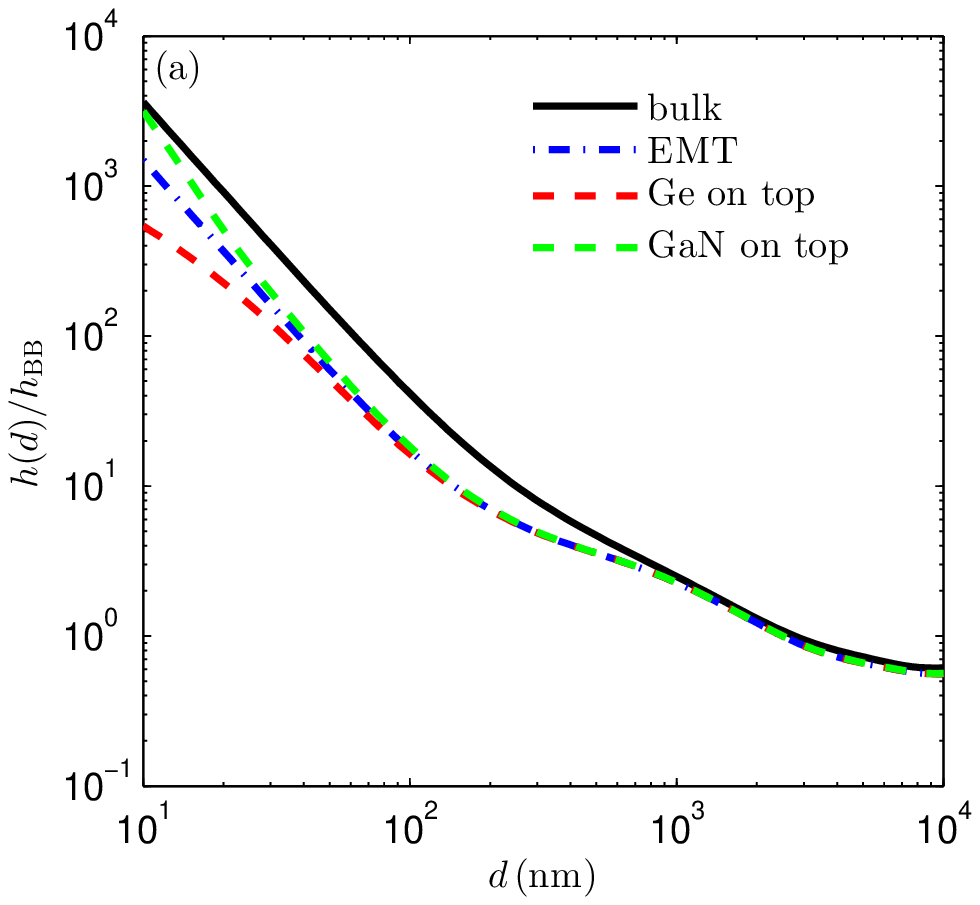}	\includegraphics[width=0.45\textwidth]{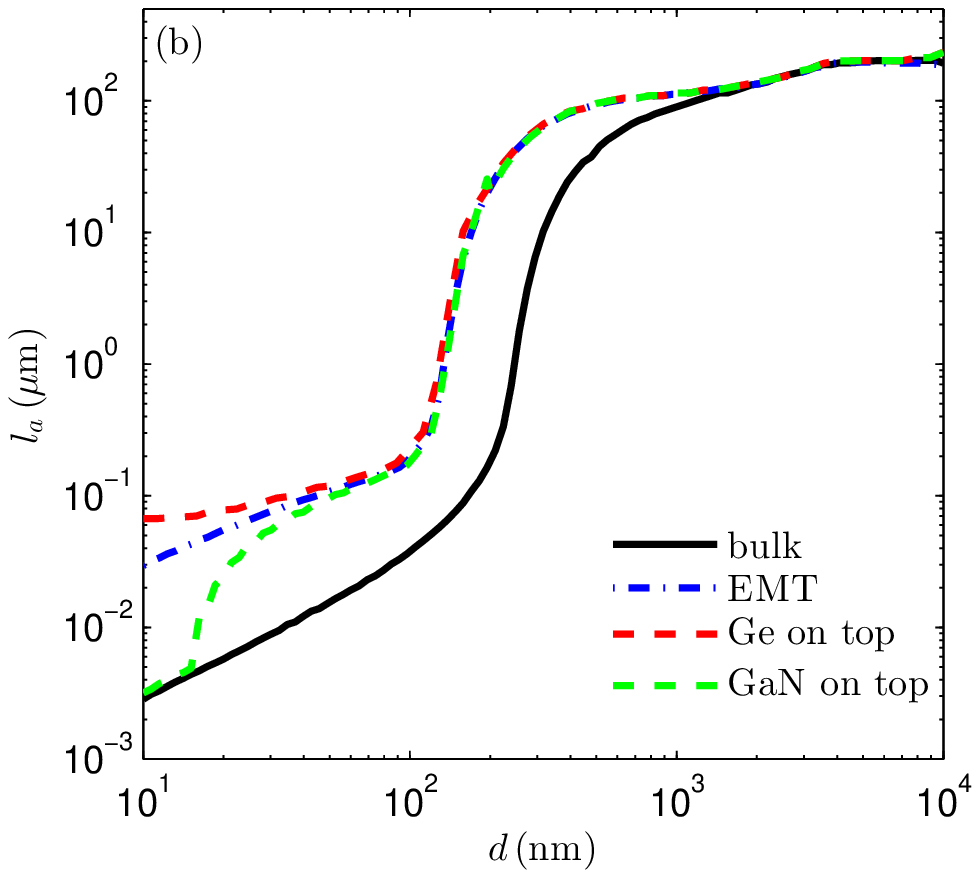} 
		\includegraphics[width=0.45\textwidth]{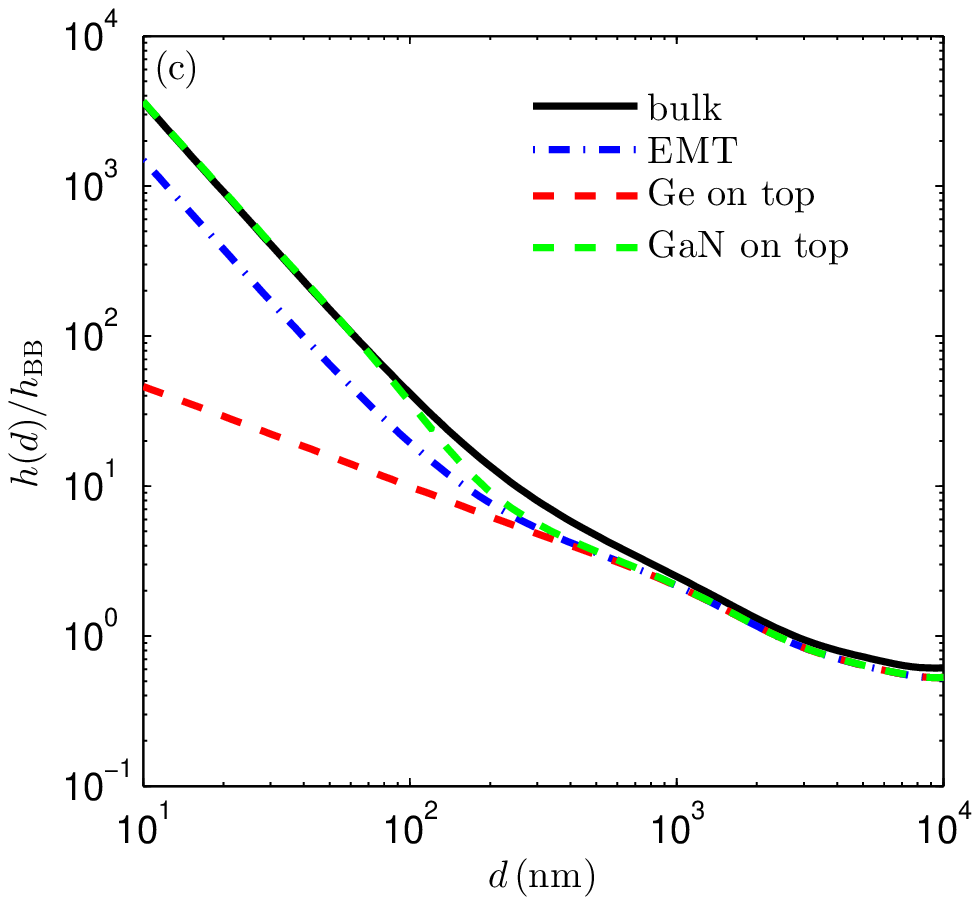}	\includegraphics[width=0.45\textwidth]{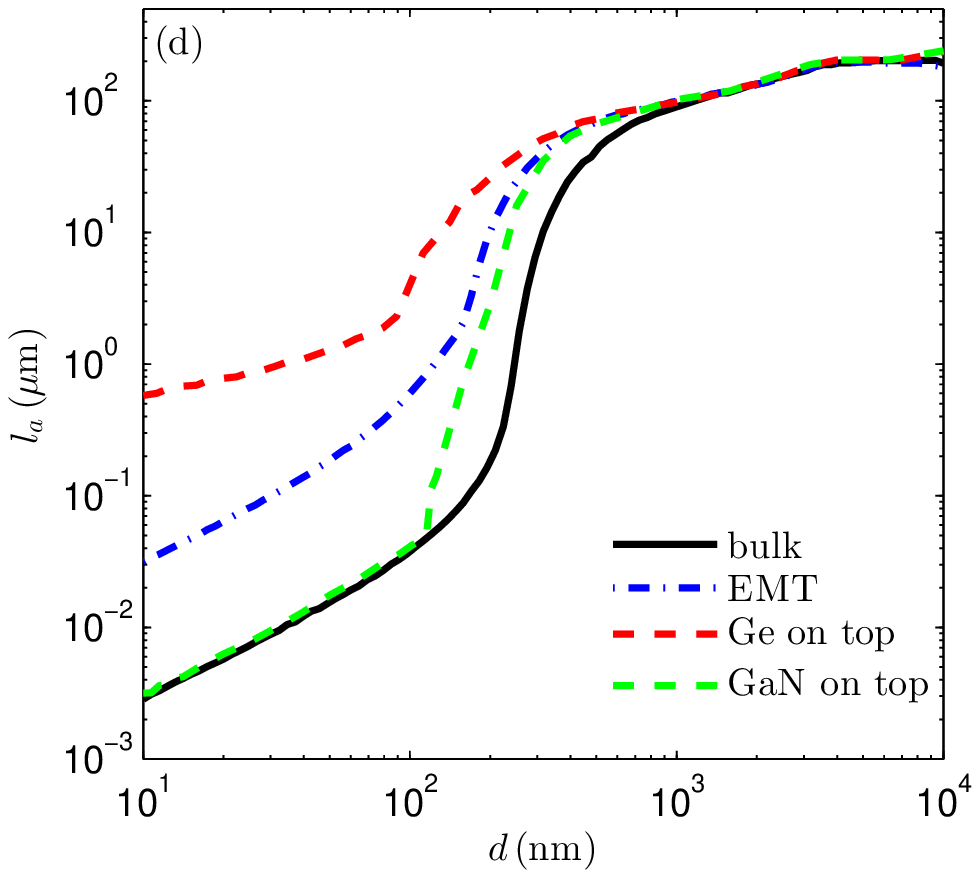} 
  \end{center}
  \caption{(a) and (c) show the total heat transfer coefficient $h(d)$ normalized to the black body value $h_{\rm BB} = 6.1\,{\rm W}{\rm m}^{-2}{\rm K}^{-1}$ for $\Lambda = 10\,{\rm nm}$ (top) and $\Lambda = 100\,{\rm nm}$ (bottom) as a function of distance $d$. (b) and (d) show total attenuation length $l_\ra$. }
	\label{Fig:HTC}
\end{figure*}

Now, let us discuss the total heat transfer coefficient $h(d)$ for the different materials. In Fig.~\ref{Fig:HTC} we show $h$ for all cases (i)-(iv) for  $\Lambda = 100\,{\rm nm}$ and $\Lambda = 10\,{\rm nm}$. First it can be seen, that the heat transfer coefficient for the hyperbolic multilayer structure (iv) with GaN as topmost layer gives the same value as bulk GaN (i) for distances smaller than the thickness of the topmost layer as can be expected~\cite{Volokitin2001,Biehs2007}. Furthermore, it can be seen that the heat transfer coefficient for the hyperbolic multilayer structure (iii) with Ge as topmost layer starts to saturate at distances smaller than the thickness of the topmost layer as found in ~\cite{BiehsEtAl2013}. The effective medium result (ii) is between the two different hyperbolic structures (iii) and (iv) and tends to overestimate the heat flux given by the hyperbolic structure (ii) with Ge as topmost layer~\cite{LiuShen2013,BiehsEtAl2013}. For more details on the applicability of effective medium theory we refer to Refs.~\cite{TschikinEtAl2013,LiuEtAl2014}.

In order to quantify the surface volume in which the most part of the incoming thermal radiation is absorbed, we determine the total attenuation length $l_\ra$ for different thicknesses $d$ of the vacuum gaps. As can be seen in Fig.~\ref{Fig:HTC}, for all shown distances the hyperbolic structures (ii)-(iv) have in general larger penetration depths than the GaN halfspace. Only for distances smaller than the thickness of the topmost layer the result for (iv) with GaN on top coincides with the result of (i), since in this case the heat flux is solely given by the surface modes of the topmost layer. To be more precise, in Figs.~\ref{Fig:HTC}(b) and \ref{Fig:HTC}(d) it can be seen that the result of (iv) coincides with (i) when $d < \Lambda$ which means when the coupling between the layers is negligible. On the other hand the effective hyperbolic structure (ii) has for distances in the near-field regime an attenuation length which is about one order of magnitude larger than that of structure (i). Finally, the attenuation length $l_\ra$ of structure (iii) with Ge on top can be even larger than the result predicted by the effective medium theory. This can be easily explained by the fact that the attenuation length is in this case at least $\Lambda/2$ since the damping inside the first Ge layer is negligible. However, we find a minimal attenuation length of about $6 \Lambda$. Hence, the total attenuation length inside the hyperbolic material (ii) can be two orders of magnitude larger than for bulk GaN. 

Note, that there is a trade-off between large heat transfer coefficients and large attenuation lengths in the near-field regime. 
At least in the strong evanescent regime where $\kappa$ is so large that 
$\gamma_\rp = \sqrt{\omega^2/c^2 \epsilon_\parallel - \kappa^2 \epsilon_\parallel/\epsilon_\perp} \approx \ri \kappa \sqrt{\epsilon_\parallel/\epsilon_\perp}$
this can be easily understood. In this case the attenuation length $L_\ra$ for a mode $(\omega,\kappa)$ is $1/2 \Im(\gamma_\rp)$. For a lossless hyperbolic material with
$\epsilon_\parallel \in \mathds{R}$ and $\epsilon_\perp \in \mathds{R}$ the attenuation length would be infinite. In contrast, for a lossy medium $L_\ra$ is finite and $L_\ra \propto 1/\kappa$.
Now, the heat flux is in this regime typically increased by increasing the number of contributing modes~\cite{BiehsEtAl2010}. This means that the heat flux is increased
by the contribution of modes with larger $\kappa$. Apparently, these modes will have a smaller attenuation length, since  $L_\ra \propto 1/\kappa$.

%%%%%%%%%%%%%%%%%%%%%%%%%%%%%%%%%%%%%%%%%%%%%%%%%%%%%%%%%%%%
\subsection{Results and discussion - infinite doublelayer structure}
%%%%%%%%%%%%%%%%%%%%%%%%%%%%%%%%%%%%%%%%%%%%%%%%%%%%%%%%%%%%

\begin{figure*}[htb]
  \begin{center}
		\includegraphics[width=0.45\textwidth]{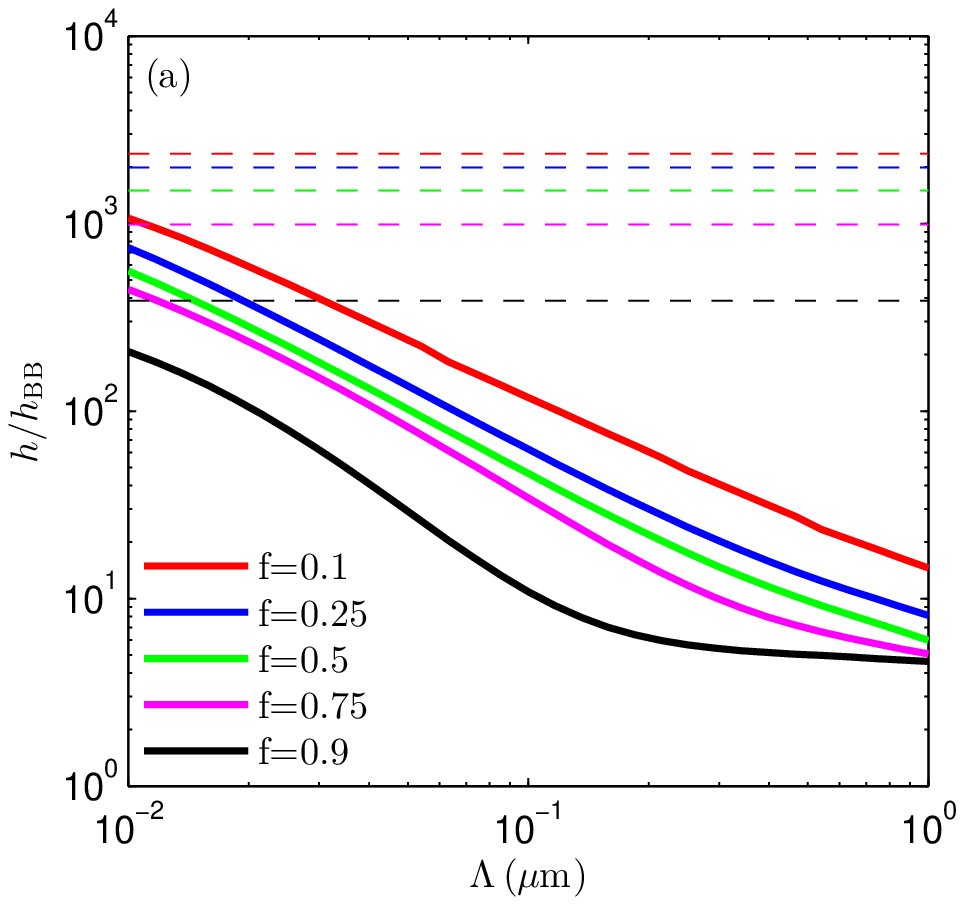}	\includegraphics[width=0.45\textwidth]{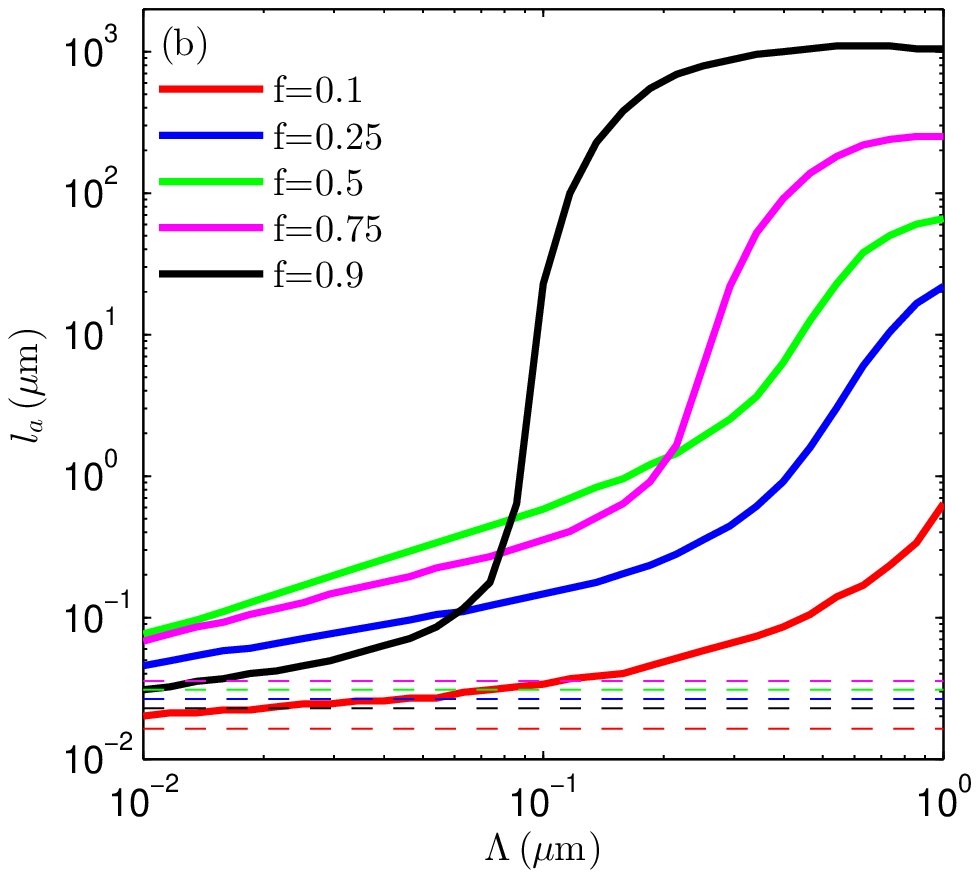} 
		\includegraphics[width=0.45\textwidth]{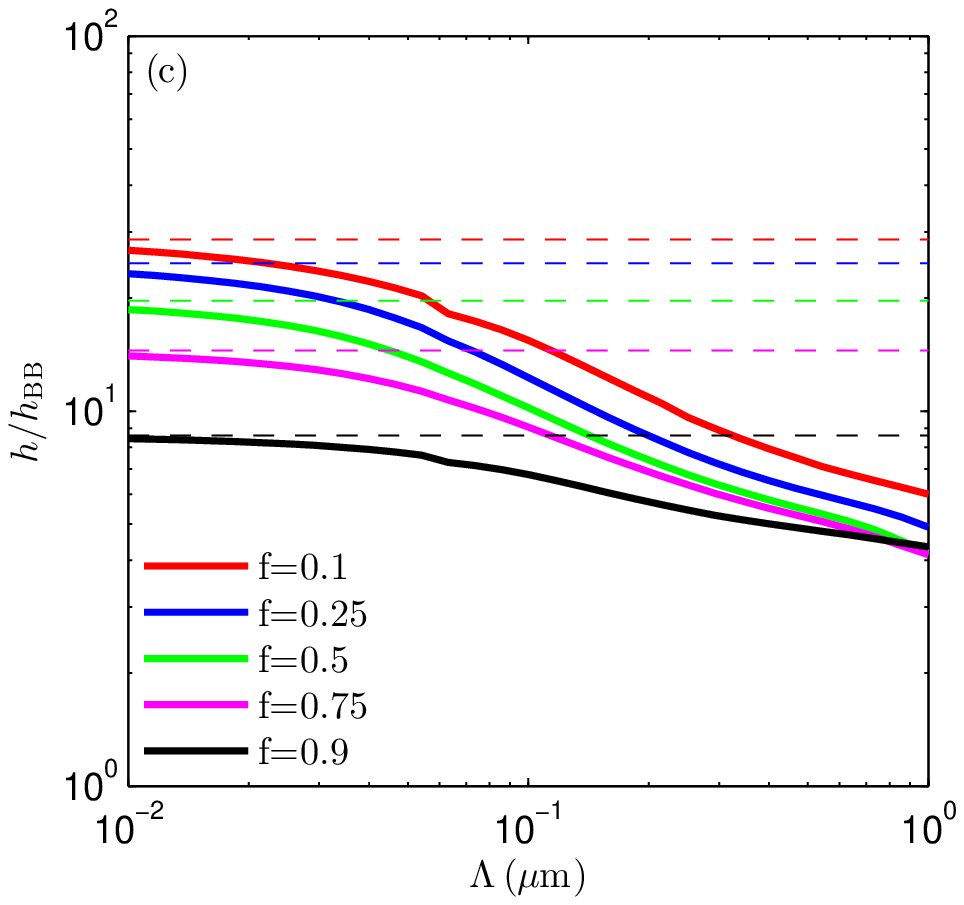}	\includegraphics[width=0.45\textwidth]{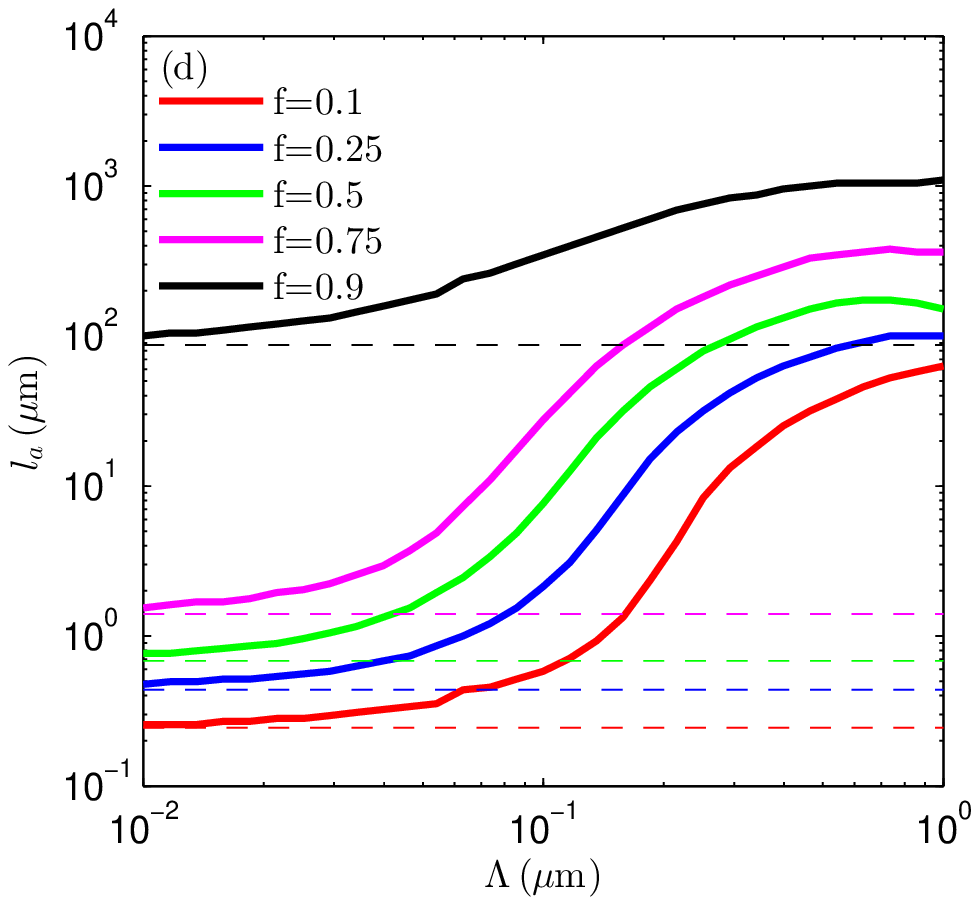} 
  \end{center}
  \caption{(a) and (c) show the total heat transfer coefficient $h$ at $z=d$ normalized to the black body value $h_{\rm BB} = 6.1\,{\rm W}{\rm m}^{-2}{\rm K}^{-1}$ for $d = 10\,{\rm nm}$ (top) and $d = 100\,{\rm nm}$ (bottom) for different filling fractions of Ge ($f=f_{\rm Ge}$) as a function of period $\Lambda$. (b) and (d) show total attenuation length $l_\ra$. The dashed lines mark the effective results.}
	\label{Fig:7}
\end{figure*}

The results discussed above are restricted to a filling fraction of $f=0.5$ and periods of $\Lambda=10\,$nm and $\Lambda=100\,$nm. Furthermore, the investigation was given for a finite periodic doublelayer structure. In order to investigate the dependence of the attenuation length $l_\ra$ as a function of the filling fraction $f$ and of the period $\Lambda$ without any spurious effects due to the finite size of the periodic structure, we consider now an infinitely extended GaN/Ge-doublelayer structure for the receiver assuming that the attenuation inside the material is determined by the Bloch wavevector (see also \cite{Slawa2014}). In this case the transmission coefficient of Eq. (\ref{EQ:TC}) simplifies for $ z\geq d$ to
\begin{equation}
  \mathfrak{T}^{({j})}(\omega,\kappa;z)= \begin{cases}
     {\rm e}^{-2\Im(K)(z-d)}(1-|r^{10}_{j}|^2)(1-|r^{12}_{j}|^2)/|D_j^{02}|^2, & \kappa < k_1\\
     4{\rm e}^{-2\Im(K)(z-d)} \, \Im(r_j^{10})\Im(r_j^{12})\mathrm e^{-2|\gamma_1|d}/|D_j^{02}|^2, & \kappa > k_1
  \end{cases}
  \label{EQ:TC0}
\end{equation}
with $D_j^{02}=1-r_j^{10}r_j^{12}{\rm e}^{2{\rm i}\gamma_1 d}$. Here $r_j^{10}$ denotes the Fresnel reflection coefficient at the surface between vacuum and the emitting halfspace on the left and $r_j^{12}$ is the reflection coefficient of the doublelayer material, which can be calculated by using the Bloch ansatz~\cite{Yeh,TschikinPC2012}. Note that $K$ in Eq.~(\ref{EQ:TC0}) denotes the Bloch wavenumber. We have checked that this approximate expressions converges to the exact result for $d \ll \Lambda$ and to the effective result for $d \gg \Lambda$.

Using this formalism, we have determined the heat transfer coefficient $h$ between a GaN bulk emitter and a doublelayer Ge/GaN structure as well as the attenuation length $l_{\rm a}$ inside the doublelayer structure for different filling fractions $f$ as a function of period $\Lambda$ for $d = 10\,{\rm nm}$ and $d = 100\,{\rm nm}$. We consider here the case where Ge is the topmost layer only. In order to compare the Bloch results with effective calculations we show in Fig.~\ref{Fig:7} also the effective results as dashed lines. It can be seen that for decreasing $\Lambda$ the Bloch curves converge to the effective results as can be expected. Furthermore, for increasing period $\Lambda$ the attenuation length $l_\ra$ increases whereas the heat transfer coefficient $h$ decreases. This tendency was already observed in Fig.~\ref{Fig:HTC}. The attenuation length depends very strongly on the filling fraction $f$ as can be seen in Figs.~\ref{Fig:7}(b) and \ref{Fig:7}(d). For the considered $\Lambda$ range the attenuation length increases for increasing $f$. In contrast to this $h$ decreases with increasing $f$. While the variation of $h$ for a fixed $f$ is in the range of two orders of magnitude for $d = 10\,{\rm nm}$ and one order of magnitude for $d=100\,$nm, the variation of $l_\ra$ is between four and three orders of magnitude, respectively. For $f=0.9$ and  $\Lambda=1\,\mu{\rm m}$ one can even reach attenuation lengths which are more than two orders of magnitude larger than the thermal wavelength. But in this case $h$ is only four times larger $h_{\rm BB}$ and the hyperbolic bands are narrow, i.e. $h$ is dominated mainly by non-hyperbolic frustrated modes explaining such large $l_\ra$ and such small $h$. On the other hand, for $f = 0.5$, i.e.\ when the hyperbolic band has its maximal width, we find for $\Lambda=1\,\mu{\rm m}$ an attenuation length of about ten times the thermal wavelength and a slightly larger heat flux than for $f = 0.9$. In general there seems to be a tendency of opposite trends for $h$ and $l_{\rm a}$ when changing $d$, $f$ or $\Lambda$. But this is only a tendency. As can be seen in Fig.~\ref{Fig:7}(b) the attenuation length changes non-monotonically when increasing the filling fraction for $\Lambda < 100\,{\rm nm}$. In this case the largest penetration depth is obtained for $f = 0.5$. Hence, depending on the purpose of the structure (large attenuation lengths or large heat fluxes or a compromise between both) one has to find the optimum of the parameters $d$, $f$ and $\Lambda$.

%%%%%%%%%%%%%%%%%%%%%%%%%%%%%%%%%%%%%%%%%%%%%%%%%%%%%%%%%%%%
\section{Conclusion}
%%%%%%%%%%%%%%%%%%%%%%%%%%%%%%%%%%%%%%%%%%%%%%%%%%%%%%%%%%%%

We have presented an exact formalism to determine the attenuation length of Poynting vector due to near- and far-field thermal radiation inside any kind of multilayer structure. In particular, we have studied the attenuation length inside multilayer hyperbolic structures composed of materials which support surface waves showing that the heat flux can be transported at longer distances than in its bulk constituents. We have shown for multilayer materials of finite size that the attenuation length inside the investigated hyperbolic structure is about one to two orders of magnitude larger than inside the bulk materials but it highly depends on the choice of the topmost layer material which can strongly screen the heat flux. Additionally, we have studied the dependence of the heat flux and the attenuation length on the filling fraction and the period for an infinite multilayer structure. It turns out that the attenuation length can be modulated by up to four orders of magnitude by changing the filling fraction and/or period of the structure. It can even be 100 times larger than the thermal wavelength for configurations where the heat flux is not dominated by the hyperbolic modes but rather by usual frustrated modes. However, there is a tendency of opposite trends for the thermal heat flux and the attenuation length, i.e. large attenuation lengths are accompanied by relatively small heat fluxes and vice versa. The long range heat transport could be advantageous for several near-field technologies. In particular, it could be used to overcome the tricky problem of the saturation in hole-electron pairs close to the surface in nTPV devices. It could also be exploited  to develop efficient heat removal systems which are able to extract the huge density of energy confined at the surface of hot bodies.

%%%%%%%%%%%%%%%%%%%%%%%%%%%%%%%%%%%%%%%%%%%%%%%%%%%%%%%%%%%%
%%%%%%%%%%%%%%%%%%%%%%%  Appendix  %%%%%%%%%%%%%%%%%%%%%%%%%
%%%%%%%%%%%%%%%%%%%%%%%%%%%%%%%%%%%%%%%%%%%%%%%%%%%%%%%%%%%%

%\section*{Appendix}
\appendix
 
\section{T and S matrix}	\label{AppendixSmatrix}

To calculate the Poynting vector in the $n$-th layer we need the respective amplitudes $a_n^{(j)}$ and $b_n^{(j)}$. In order to determine these amplitudes we make the following two steps. In a first step we employ the usual continuity conditions for the Green's tensors at the interfaces $z=z_n$ to get the transfer matrix
\begin{equation}
	\begin{pmatrix} a_n  \\ b_n  \end{pmatrix} = \begin{pmatrix} 
T_1 & T_2 \\ T_3 & T_4
\end{pmatrix}  \begin{pmatrix} a_{n+1} \\ b_{n+1}  \end{pmatrix}
	\label{EQ:Tmatrix}
\end{equation}
for both s- and p- polarized modes connecting the amplitudes of adjacent layers. The matrix elements of the T matrix are given by
\begin{alignat}{4}
  T_1 &= \frac{1}{t_{n,n+1}}{\rm e}^{-{\rm i} (\gamma_n-\gamma_{n+1}) z_n}, 
  &T_2 &= \frac{r_{n,n+1}}{t_{n,n+1}}{\rm e}^{-{\rm i} (\gamma_n+\gamma_{n+1}) z_n}, \\
  T_3 &= \frac{r_{n,n+1}}{t_{n,n+1}}{\rm e}^{ {\rm i} (\gamma_n+\gamma_{n+1}) z_n}, 
  &T_4 &= \frac{1}{t_{n,n+1}}{\rm e}^{{\rm i} (\gamma_n-\gamma_{n+1}) z_n},      
\end{alignat}
where $r_{i,j}$ and $t_{i,j}$ are the Fresnel reflection and transmission coefficients for s- and p-polarized modes
\begin{alignat}{4}
  r_{i,j}^{(\rm s)}&=\frac{\gamma_i -\gamma_j}{\gamma_i + \gamma_j},\quad
  &t_{i,j}^{(\rm s)}&=\frac{2\gamma_i}{\gamma_i + \gamma_j},  \\
  r_{i,j}^{(\rm p)}&=\frac{\epsilon_j\gamma_i -\epsilon_i\gamma_j}{\epsilon_j\gamma_i + \epsilon_i\gamma_j}, \quad 
  &t_{i,j}^{(\rm p)}&=\frac{2\sqrt{\epsilon_i\epsilon_j}\gamma_i}{\epsilon_j\gamma_i + \epsilon_i\gamma_j}. 
\end{alignat}
For computational reasons, in a second step we determine the scattering matrix connecting the amplitudes of the incoming and outgoing waves~\cite{AuslenderHava1996,Francoeur2009}
\begin{equation}
  \begin{pmatrix} a_n  \\ b_0  \end{pmatrix} = \begin{pmatrix} 
    S_1(n) & S_2(n) \\ S_3(n) & S_4(n) 
   \end{pmatrix}  \begin{pmatrix} a_0 \\ b_n  \end{pmatrix}
	\label{EQ:Smatrix}
\end{equation}
with $S_1(0)=1$, $S_2(0)=0$, $S_3(0)=0$, and $S_4(0)=1$. Using the T matrix in Eq.~(\ref{EQ:Tmatrix}) and the S matrix for the $n$-th layer~(\ref{EQ:Smatrix}) we can determine the S matrix for the $(n+1)$-th layer
\begin{equation}
	\begin{pmatrix} a_{n+1}  \\ b_0  \end{pmatrix} = \begin{pmatrix} 
     S_1(n+1) & S_2(n+1) \\ S_3(n+1) & S_4(n+1) 
	\end{pmatrix}  \begin{pmatrix} a_0 \\ b_{n+1}  \end{pmatrix}
	\label{EQ:Smatrix2}
\end{equation}
with the $(n+1)$-th S-matrix elements given by
\begin{alignat}{4}
  S_1(n+1)&= \frac{S_1(n)}{{T_1 - S_2(n)T_3}} \qquad 
  &S_2(n+1)&= \frac{{S_2(n)T_4 - T_2}}{{T_1-S_2(n)T_3}}, \\
  S_3(n+1)&= S_3(n) + S_4(n)T_3 S_1(n+1), \qquad
  &S_4(n+1)&= S_4(n)T_3S_2(n+1) + S_4(n)T_4.
\end{alignat}
From the condition $a_0=1$ and $b_{N+1}=0$ we can determine all other amplitudes by means of the S-matrix method.

%%%%%%%%%%%%%%%%%%%%%%%%%%%%%%%%%%%%%%%%%%%%%%%%%%%%%%%%%%%%
%%%%%%%%%%%%%%%%%%%%  Acknowledgments  %%%%%%%%%%%%%%%%%%%%%
%%%%%%%%%%%%%%%%%%%%%%%%%%%%%%%%%%%%%%%%%%%%%%%%%%%%%%%%%%%%

\section*{Acknowledgments}

M.\ T.\ gratefully acknowledges support from the Stiftung der Metallindustrie im Nord-Westen. S.-A.\ B., M.\ T.\ and P.\ B.-A.\ acknowledge financial support by the DAAD and Partenariat Hubert Curien Procope Program (project 55923991). The authors from Hamburg University of Technology gratefully acknowledge financial support from the German Research Foundation (DFG) via SFB 986 "M$^3$", project C1.




\begin{thebibliography}{00}


%Heat flux
  \bibitem{PvH1971} D. Polder and M. van Hove, ``Theory of Radiative Heat Transfer between Closely Spaced Bodies,'' Phys. Rev. B {\bf 4}, 3303 (1971).
%experiment
  \bibitem{KittelEtAlPRL2005} A. Kittel, W. M\"{u}ller-Hirsch, J. Parisi, S.-A. Biehs, D. Reddig, and M. Holthaus, ``Near-field heat transfer in a scanning thermal microscope,'' Phys. Rev. Lett. {\bf 95}, 224301 (2005). 
  \bibitem{NanolettArvind} A. Narayanaswamy, S. Shen, and G. Chen, ``Near-field radiative heat transfer between a sphere and a substrate,'' Phys. Rev. B {\bf 78}, 115303 (2008).
  \bibitem{NatureEmmanuel} E. Rousseau, A. Siria, G. Jourdan, S. Volz, F. Comin, J. Chevrier, and J.-J. Greffet, ``Radiative heat transfer at the nanoscale,''  Nat. Photonics {\bf 3}, 514 (2009).
  \bibitem{HuEtAl2008} L. Hu, A. Narayanaswamy, X. Chen, and G. Chen, ``Near-field thermal radiation between two closely spaced glass plates exceeding Planck's blackbody radiation law,'' App. Phys. Lett. {\bf 92}, 133106 (2008).
  \bibitem{ShenEtAl2009} S. Shen, A. Narayanaswamy, and G. Chen, ``Surface Phonon Polaritons Mediated Energy Transfer between Nanoscale Gaps,'' Nano Lett. {\bf 9}, 2909 (2009).
  \bibitem{ShiEtAl2013} J. Shi, P. Li, B. Liu, and S. Shen, ``Tuning near field radiation by doped silicon,'' App. Phys. Lett. {\bf 102}, 183114 (2013).
  \bibitem{Ottens2011} R. S. Ottens, V. Quetschke, S. Wise, A. A. Alemi, R. Lundock, G. Mueller, D. H. Reitze, D. B. Tanner, and B. F. Whiting, ``Near-Field Radiative Heat Transfer between Macroscopic Planar Surfaces,'' Phys. Rev. Lett. {\bf 107}, 014301 (2011).
  \bibitem{Kralik2012} T. Kralik, P. Hanzelka, M. Zobac, V. Musilova, T. Fort, and M. Horak, ``Strong Near-Field Enhancement of Radiative Heat Transfer between Metallic Surfaces,'' Phys. Rev. Lett. {\bf 109}, 224302 (2012).
%sphere plane 
  \bibitem{Otey2011} C. Otey and S. Fan, ``Numerically exact calculation of electromagnetic heat transfer between a dielectric sphere and plate,'' Phys. Rev. B {\bf 84}, 245431 (2011).
  \bibitem{KruegerEtAl2011} M. Kr\"{u}ger, T. Emig, and M. Kardar, ``Nonequilibrium Electromagnetic Fluctuations: Heat Transfer and Interactions,'' Phys. Rev. Lett. {\bf 106}, 210404 (2011).
%thermal imaging
  \bibitem{Yannick} Y. De Wilde, F. Formanek, R. Carminati, B. Gralak, P.-A. Lemoine, K. Joulain, J.-P. Mulet, Y. Chen, and J.-J. Greffet, ``Thermal radiation scanning tunnelling microscopy,''  Nature {\bf 444}, 740 (2006).
  \bibitem{KittelEtAl2008} A. Kittel , U. Wischnath , J. Welker , O. Huth , F. R\"{u}ting, and S.-A. Biehs, ``Near-field thermal imaging of nano-structured surfaces,'' Appl. Phys. Lett. {\bf 93}, 193109 (2008).
  \bibitem{HuthEtAl2011} F. Huth, M. Schnell, J. Wittborn, N. Ocelic, and R. Hillenbrand, ``Infrared-spectroscopic nanoimaging with a thermal source,'' Nat. Mater. {\bf 10}, 352 (2011).
  \bibitem{WorbesEtAl2013} L. Worbes, D. Hellmann, and A. Kittel, ``Enhanced Near-Field Heat Flow of a Monolayer Dielectric Island,'' Phys. Rev. Lett. {\bf 110}, 134302 (2013).
%Thermal management  
  \bibitem{OteyEtAl2010} C. R. Otey, W. T. Lau, and S. Fan, ``Thermal Rectification through Vacuum,'' Phys. Rev. Lett. {\bf 104}, 154301 (2010).
  \bibitem{IizukaEtAl2012} H. Iizuka and S. Fan, ``Rectification of evanescent heat transfer between dielectric-coated and uncoated silicon carbide plates,'' J. Appl. Phys. {\bf 112}, 024304 (2012).
  \bibitem{HuangEtAl2013} J. Huang, Q. Li, Z. Zheng, and Y. Xuan, ``Thermal rectification based on thermochromic materials,'' Int. J. Heat Mass Transfer {\bf 67}, 575 (2013).
  \bibitem{PBASAB_PRL2014} P. Ben-Abdallah and S.-A. Biehs, ``Near-feld thermal transistor,'' Phys. Rev. Lett. {\bf 112}, 044301 (2014). 
  \bibitem{PBASABDiode2013} P. Ben-Abdallah and S.-A. Biehs, ``Phase-change radiative thermal diode,'' Appl. Phys. Lett. {\bf 103}, 191907 (2013). 
  \bibitem{Nefzaoui2013} E. Nefzaoui, J. Drevillon, Y. Ezzahri, and K. Joulain, ``Simple far-field radiative thermal rectifier using Fabry–Perot cavities based infrared selective emitters,'' Applied Optics {\bf 53}, 3479 (2014).
% Storage
  \bibitem{KubitskyiEtAl2014} V. Kubytskyi, S.-A. Biehs, P. Ben-Abdallah, ``Radiative Bistability and Thermal Memory,'' Phys. Rev. Lett. {\bf 113}, 074301 (2014). 
  \bibitem{DyakovEtAl2014} S. A. Dyakov, J. Dai, M. Yan, M. Qiu, ``Near field thermal memory device,'' e-print arXiv:1408.5831. 
% TPV
  \bibitem{MatteoEtAl2001} R. S. DiMatteo, P. Greiff, S. L. Finberg, K. A. Young-Waithe, H. K. Choy, M. M. Masaki, and C. G. Fonstad, ``Enhanced photogeneration of carriers in a semiconductor via coupling across a nonisothermal nanoscale vacuum gap,'' Appl. Phys. Lett. {\bf 79}, 1894 (2001).
  \bibitem{NarayanaswamyChen2003} A. Narayanaswamy and G. Chen, ``Surface modes for near field thermophotovoltaics,'' Appl. Phys. Lett. {\bf 82}, 3544 (2003).
  \bibitem{LarocheEtAl06} M. Laroche, R. Carminati, and J.-J. Greffet, ``Near-field thermophotovoltaic energy conversion,'' J. Appl. Phys. {\bf 100}, 063704 (2006).
  \bibitem{ParkEtAl2007} K. Park, S. Basu, W. P. King, and Z. M. Zhang, ``Performance analysis of near-field thermophotovoltaic devices considering absorption distribution,'' J. Quant. Spect. Rad. Transf. {\bf 109}, 305 (2008).
  \bibitem{ZhangReview} S. Basu, Z. M. Zhang, and C. J. Fu, ``Review of near-field thermal radiation and its application to energy conversion,'' Int. J. Energy Res. {\bf 33}, 1203 (2009).
  \bibitem{GuoEtAl2013} Y. Guo and Z. Jacob, ``Thermal hyperbolic metamaterials,'' Opt. Express {\bf 21}, 15014 (2013).
%surface modes
  \bibitem{JoulainEtAl2005} K. Joulain, J.-P. Mulet, F. Marquier, R. Carminati, and J.-J. Greffet, ``Surface electromagnetic waves thermally excited: Radiative heat transfer, coherence properties and Casimir forces revisited in the near field,'' Surf. Sci. Rep. {\bf 57}, 59 (2005).
  \bibitem{BiehsEtAl2010} S.-A. Biehs, E. Rousseau, and J.-J. Greffet, ``A mesoscopic description of radiative heat transfer at the nanoscale,'' Phys. Rev. Lett. {\bf 105}, 234301 (2010). 
%limits 
  \bibitem{PBAjoulain2010} P. Ben-Abdallah and K. Joulain, ``Fundamental limits for noncontact transfers between two bodies,'' Phys. Rev. B {\bf 82}, 121419(R) (2010).
  \bibitem{Volok2004} A. I. Volokitin and B. N. J. Persson, ``Resonant photon tunneling enhancement of the radiative heat transfer,'' Phys. Rev. B {\bf 69}, 045417 (2004).
  \bibitem{BasuZhanglim2009} S. Basu and Z. M. Zhang, ``Maximum energy transfer in near-field thermal radiation at nanometer distances,'' J. App. Phys. {\bf 105}, 093535 (2009).
  \bibitem{WangEtAl2009} X. J. Wang, S. Basu, and Z. M. Zhang, ``Parametric optimization of dielectric functions for maximizing nanoscale radiative transfer,'' J. Phys. D: Appl. Phys. {\bf 42}, 245403 (2009).
%PD for surface modes
  \bibitem{BasuZhang2009} S. Basu and Z. M. Zhang, ``Ultrasmall penetration depth in nanoscale thermal radiation,'' Appl. Phys. Lett. {\bf 95}, 133104 (2009).
  \bibitem{BasuZhang2011} S. Basu and M. Francoeur, ``Penetration depth in near-field radiative heat transfer between metamaterials,'' Appl. Phys. Lett. {\bf 99}, 143107 (2011).
%HF hyperbolic modes
  \bibitem{BiehsEtAl2012} S.-A. Biehs, M. Tschikin, and P. Ben-Abdallah, ``Hyperbolic Metamaterials as an Analog of a Blackbody in the Near Field,'' Phys. Rev. Lett. {\bf 109}, 104301 (2012).
  \bibitem{GuoEtAl2012} Y. Guo, C. L. Cortes, S. Molesky, and Z. Jacob, ``Broadband super-Planckian thermal emission from hyperbolic metamaterials,'' Appl. Phys. Lett. {\bf 101}, 131106 (2012).
%HMM
  \bibitem{SmithSchurig2003} D. R. Smith and D. Schurig, ``Electromagnetic Wave Propagation in Media with Indefinite Permittivity and Permeability Tensors,'' Phys. Rev. Lett. {\bf 90}, 077405 (2003).
% hyperbolic materials in nature
  \bibitem{DrachevEtAl2013} V. P. Drachev, V. A. Podolskiy, and A. V. Kildishev,``Hyperbolic metamaterials: new physics behind a classical problem,'' Opt. Express {\bf 21}, 15048 (2013).
  \bibitem{ThompsonEtAl1998} D. W. Thompson, M. J. DeVries, T. E. Tiwald, and J. A. Woollam, `` Determination of optical anisotropy in calcite from ultraviolet to mid-infrared by generalized ellipsometry,'' Thin Solid Films {\bf 313-314}, 341 (1998).
  \bibitem{CaldwellEtAl2014} J. D. Caldwell, A. Kretinin, Y. Chen, V. Giannini, M. M. Fogler, Y. Francescato, C. T. Ellis, J. G. Tischler, C. R. Woods, A. J. Giles, M. Hong, K. Watanabe, T. Taniguchi, S. A. Maier, K. S. Novoselov, ``Sub-diffractional, volume-confined polaritons in a natural hyperbolic material: hexagonal boron nitride,'' e-print arXiv:1404.0494.
%multilayer HMM
  \bibitem{HoffmanEtAl2007} A. J. Hoffman, L. Alekseyev, S. S. Howard, K. J. Franz, D. Wasserman, V. A. Podolskiy, E. E. Narimanov, D. L. Sivco, and C. Gmachl, ``Negative refraction in semiconductor metamaterials,'' Nat. Mater. {\bf 6}, 946 (2007).  
  \bibitem{KrishnamoorthyEtAl2012} H. N. S. Krishnamoorthy, Z. Jacob, E. Narimanov, I. Kretzschmar, and V. M. Menon, ``Topological Transitions in Metamaterials,'' Science {\bf 336}, 205 (2012). 
  \bibitem{LangEtAl2013} S. Lang, H. S. Lee, A. Y. Petrov, M. St\"{o}rmer, M. Ritter, and M. Eich, ``Gold-silicon metamaterial with hyperbolic transition in near infrared,'' Appl. Phys. Lett. {\bf 103}, 21905 (2013).
%wire hyperbolic materials   
  \bibitem{WangbergEtAl2006} R. Wangberg, J. Elser, E. E. Narimanov, and V. A. Podolskiy, ``Nonmagnetic nanocomposites for optical and infrared negative-refractive-index media,'' J. Opt. Soc. Am. B {\bf 23}, 498 (2006). 
  \bibitem{YaoEtAl2008} J. Yao, Z. Liu, Y. Liu, Y. Wang, C. Sun, G. Bartal, A. M. Stacy, and X. Zhang, ``Optical Negative Refraction in Bulk Metamaterials of Nanowires,'' Science {\bf 321}, 930 (2008).
  \bibitem{NoginovEtAl2009} M. A. Noginov, Y. A. Barnakov, G. Zhu, T. Tumkur, H. Li, and E. E. Narimanov, ``Bulk photonic metamaterial with hyperbolic dispersion,'' Appl. Phys. Lett. {\bf 94}, 151105 (2009).
  \bibitem{CaiEtAl2010} W. Cai and V. Shalaev, {\em Optical Metamaterials: Fundamentals and Applications} (Springer, New York, Dordrecht, Heidelberg, London, 2010).
%  \bibitem{DrachevEtAl2013} V. P. Drachev, V. A. Podolskiy, and A. V. Kildishev, ``Hyperbolic metamaterials: new physics behind a classical problem,'' Opt. Express {\bf 21}, 15048 (2013).
  \bibitem{PoddubnyEtAl2013} A. Poddubny, I. Iorsh, P. Belov, and Y. Kivshar, ``Hyperbolic metamaterials,'' Nat. Photonics {\bf 7}, 948 (2013).
%hyperbolic tpv
  \bibitem{Nefedov2011} I. S. Nefedov and C. R. Simovski, ``Giant radiation heat transfer through micron gaps,'' Phys. Rev. B {\bf 84}, 195459 (2011).
  \bibitem{Simovski2013} C. Simovski, S. Maslovski, I. Nefedov, and S. Tretyakov, ``Optimization of radiative heat transfer in hyperbolic metamaterials for thermophotovoltaic applications,'' Opt. Express {\bf 21}, 14988 (2013).
%PD in HMMs
	\bibitem{Slawa2014} S. Lang, M. Tschikin, S.-A. Biehs,  A. Y. Petrov, and M. Eich, ``Large penetration depth of near-field heat flux in hyperbolic media,'' Appl. Phys. Lett. {\bf 104}, 121903 (2014).
%HF hyperbolic nanowire materials, SM vs HM   
  \bibitem{LiuShen2013} B. Liu and S. Shen, ``Broadband near-field radiative thermal emitter/absorber based on hyperbolic metamaterials: Direct numerical simulation by the Wiener chaos expansion method,'' Phys. Rev. B {\bf 87}, 115403 (2013).
  \bibitem{BiehsEtAl2013} S.-A. Biehs, M. Tschikin, R. Messina, and P. Ben-Abdallah, ``Super-Planckian Near-Field Thermal Emission with Phonon-Polaritonic Hyperbolic Metamaterials,'' Appl. Phys. Lett. {\bf 102}, 131106 (2013).
%Limits of effective medium theory 
  \bibitem{TschikinEtAl2013} M. Tschikin, S.-A. Biehs, R. Messina, and P. Ben-Abdallah, ``On the limits of the effective description of hyperbolic materials in presence of surface waves,'' J. Opt. {\bf 15}, 105101 (2013). 
%
%S matrix
	\bibitem{AuslenderHava1996} M. Auslender and S. Hava, ``Scattering-matrix propagation algorithm in full-vectorial optics of multilayer grating structures,'' Opt. Lett. {\bf 21}, 1765 (1996).
  \bibitem{Francoeur2009} M. Francoeur, P. Meng\"{u}\c{c}, and R. Vaillon, ``Solution of near-field thermal radiation in one-dimensional layered media using dyadic Green's functions and the scattering matrix method,'' J. Quant. Spect. Rad. Transf. {\bf 110}, 2002 (2009).
  \bibitem{Sipe} J. E. Sipe, ``New Green-function formalism for surface optics,'' J. Opt. Soc. Am. B {\bf 4}, 481 (1987).
  \bibitem{RytovBook1989} S. M. Rytov, Y. A. Kravtsov, and V. I. Tatarskii, {\em Principles of Statistical Radiophysics 3} (Springer-Verlag, 1989).
   \bibitem{BrightEtAl2014} T. J. Bright, X. L.  Liu, and  Z.M. Zhang, ``Energy Streamlines in Near-Field Radiative Heat Transfer Between Hyperbolic Metamaterials,'' Optics Express {\bf 22}, A1112 (2014).
%FDT
  \bibitem{Kubo} R. Kubo, ``The fluctuation-dissipation theorem,'' Rep. Prog. Phys. {\bf 29}, 255 (1966).
%Quantum
  \bibitem{JanowiczEtAl2003} M. Janowicz, D. Reddig, and M. Holthaus ``Quantum approach to electromagnetic energy transfer between two dielectric bodies,'' Phys. Rev. A {\bf 68}, 043823 (2003).  
%Greens function
  \bibitem{Arvind2014} A. Narayanaswamy and Y. Zheng, ``A Green's function formalism of energy and momentum transfer in fluctuational electrodynamics,'' J. Quant. Spec. Rad. Trans. {\bf 132}, 12 (2014).
%Multilayer
  \bibitem{Yeh} P. Yeh, {\em Optical Waves in Layered Media} (Wiley, Hoboken, 2005).
%Material Properties
  \bibitem{Adachi} S. Adachi, {\em ``III-V Compound Semiconductors'' in Handbook on Physical Properties of Semiconductors} (Springer, 2004), Vol. 2.
%single layer 
 \bibitem{Volokitin2001} A. I. Volokitin and B. N. J. Persson, ``Radiative heat transfer between nanostructures,'' Phys. Rev. B {\bf 63}, 205404 (2001).
            \bibitem{Biehs2007} S.-A. Biehs, ``Thermal heat radiation, near-field energy density and near-field radiative heat transfer of coated materials,'' EPJ B {\bf 58}, 423 (2007).
  \bibitem{MillerEtAl2014} O. D. Miller, S. G. Johnson, and A. W. Rodriguez, ``Effectiveness of Thin Films in Lieu of Hyperbolic Metamaterials in the Near Field,'' Phys. Rev. Lett. {\bf 112}, 157402 (2014).
%nanowire HMM
	\bibitem{KidwaiEtAl2012} O. Kidwai, S. V. Zhukovsky, and J. E. Sipe, ``Effective-medium approach to planar multilayer hyperbolic metamaterials: Strengths and limitations,'' Phys. Rev. A {\bf 85}, 053842 (2012).
  \bibitem{LiuEtAl2014} X.L. Liu, T.J. Bright and Z.M. Zhang, ``Application Conditions of Effective Medium Theory in Near-Field Radiative Heat Transfer Between Multilayered Metamaterials,''J. Heat Trans. {\bf 136}, 092703 (2014).
%layer
	\bibitem{TschikinPC2012} M. Tschikin, P. Ben-Abdallah, and S.-A. Biehs, ``Coherent thermal conductance of 1-D photonic crystals,'' Phys. Lett. A {\bf 376}, 3462 (2012).

\end{thebibliography}


\end{document}